\documentclass[letterpaper,12pt,notoc]{JHEP3}

\usepackage{graphicx}
\usepackage{amsmath}
\usepackage{amssymb}

\preprint{ \hbox{}\hfill arXiv:1002.3760}

\title{AdS$_3$ Vacua and RG Flows in Three Dimensional Gauged
Supergravities}

\author{Edi Gava$^a$, Parinya Karndumri$^b$ and K. S. Narain$^c$\\
$^a$INFN, Sezione di Trieste, Italy\\
$^b$International School for Advanced Studies (SISSA), via Bonomea
265, 34136 Trieste, Italy \\
$^c$The Abdus Salam International Centre for Theoretical Physics,
Strada Costiera 11, 34100 Trieste, Italy \\
E-mail: \email{gava$@$ictp.it}, \email{karndumr$@$sissa.it},
\email{narain$@$ictp.it}}

\abstract{We study $AdS_3$ supersymmetric vacua in $N=4$ and $N=8$,
three dimensional gauged supergravities, with scalar
 manifolds $(\frac{SO(4,4)}{SO(4)\times
SO(4)} )^2$ and  $\frac{SO(8,8)}{SO(8)\times SO(8)} $,
non-semisimple Chern-Simons gaugings $SO(4)\ltimes {\bf R}^6$ and
$(SO(4)\ltimes {\bf R}^6)^2$, respectively. These are in turn
equivalent to  $SO(4)$ and $SO(4)\times SO(4)$ Yang-Mills theories
coupled to supergravity. For the $N=4$ case, we study
renormalization group flows between UV and IR
 $AdS_3$ vacua with the same amount
of supersymmetry: in one case, with (3,1) supersymmetry, we can find
an analytic solution whereas in another, with  (2,0) supersymmetry,
we give a numerical solution. In both cases, the flows turn out to
be v.e.v. flows, i.e. they are driven by the expectation value of a
relevant operator in the dual $SCFT_2$. These provide  examples of
v.e.v. flows between two $AdS_3$ vacua within a gauged supergravity
framework.} \keywords{AdS/CFT correspondence, gauge-gravity
correspondence}
\begin{document}
\section{Introduction}
Three dimensional gauged supergravities turn out to possess a very
rich structure, and one reason to be interested in them, apart from
their intrinsic geometrical elegance, is that they offer a
convenient arena to discuss various aspects of $AdS_3/CFT_2$
correspondence, much in the same way the study of  various
backgrounds of five-dimensional gauged supergravity has been useful
in uncovering interesting phenomena in the dual four dimensional
Yang-Mills theory\cite{fgpw, an, gir}. \\ \indent The construction
of three dimensional, $N$-extended, gauged supergravities has been
worked out systematically for any $N \leq 16$ in \cite{dewit}
extending previous results on $N=8,16$ obtained in \cite{ns1, ns3,
ns4}. When gauging isometries of the scalar manifold of the
original, ungauged supergravity theory, one introduces gauge fields
which have Chern-Simons kinetic terms and therefore do not represent
propagating degrees of freedom. On the other hand, when reducing a
higher dimensional supergravity theory down to three dimensions,
which is the instance we are interested in, one generically obtains
gauge-fields with Yang-Mills like kinetic terms. The apparent puzzle
was solved in \cite{dewit} and \cite{csym} and has to do with the
duality between gauge fields and scalars in three dimensional
space-time: more precisely,  it has been shown there that, if the
gauge group is not semisimple, but contains nilpotent shift
symmetries, i.e. it is of the form $G\ltimes {\bf R}^{{\rm dim}G}$,
then one can integrate out half of the $2\, {\rm dim}G$ Chern-Simons
gauge fields to produce a Yang-Mills action for the remaining ones.
At the same time, ${\rm dim} G$  scalars can be set to zero by using
the shift symmetries. In other words, one trades scalars with
vectors and, of course, the number of physical degrees of freedom is
unchanged. This mechanism has been employed, for example, in
\cite{ns2} for $N=8$, where it has been shown that  a gauging by
$SO(4)\ltimes {\bf R}^6$ indeed reproduces, at the $N=8$ point in
the scalar manifold, the Kaluza-Klein spectrum of the
six-dimensional (2,0) supergravity on $AdS_3\times S^3$\cite{de
boer}. The latter is the background one obtains by taking the near
horizon geometry of a D1-D5 system of type IIB theory on $K3$ or
$T^4$, corresponding to a $CFT_2$ with (4,4) supersymmetry. \\
\indent In this paper, we analyze two examples of gauged
supergravities with non-semisimple gauging, with $N=4$ and $N=8$
supersymmetry, whose scalar manifolds take the forms of
$(\frac{SO(4,4)}{SO(4)\times SO(4)})^2$ and
$\frac{SO(8,8)}{SO(8)\times SO(8)}$, respectively. As for the
gauging, we will consider gauge groups  $SO(4)\ltimes {\bf R}^6$ and
 $(SO(4)\ltimes {\bf R}^6)^2$, respectively. These turn out to
 be  subgroups of the isometry groups which can be gauged consistently
 with supersymmetry, as will be shown.
\\ \indent We will study supersymmetric $AdS_3$ vacua in both of these theories,
 with various amount of preserved supersymmetries.
 In the $N=4$ case, we will be able to study the flow between different vacua
 with different cosmological constants but the same amount of supersymmetry.
 Quite remarkably, we will be able to find an analytic flow solution between
 vacua with (3,1) supersymmetry involving two active scalar fields. For the case
 of flow between (2,0) vacua which involves three active scalars, we will discuss
 a numerical flow solution. The flows turn out to be v.e.v. flows
 driven by vacuum expectation values of some operators in the UV.
Examples of v.e.v. flows are known in four dimensional
super-conformal field theories, in particular in $N=2$ SCFT, where
they have been studied using Seiberg-Witten solution in connection
with the Argyres-Douglas fixed points\cite{ad,aw,st}. To the best of
our knowledge, these are the first examples of v.e.v. flows between
two $AdS$ vacua in a gauged supergravity context.
\\ \indent From the higher dimensional perspective, the case with $N=8$
supersymmetry (or better its maximally symmetric vacuum) is related
to the brane configuration in type IIB theory whose near horizon
geometry is $AdS_3\times S^3\times S^3\times S^1$ \cite{de boer2},
dual to a $CFT_2$  with ``large" $(4,4)$ superconformal
algebra\cite{HS1, HS2}. For the $N=4$ case, which has a (4,0)
vacuum, the ten dimensional interpretation is far  less clear. It
could be related to some warped or orbifolded versions of the
previous case. It would be interesting to establish this.\\ \indent
 The paper is organized as follows. In section 2, we review
 the features of three dimensional gauged supergravity
 in the case where the target manifold is a symmetric  space.
 In section 3, we specialize at the $N=4$ theory and describe the vacua we found.
 In section 4, we discuss the analytic flow solutions between (3,1) vacua, and numerical flow solution between (2,0) vacua.
 In section 5, we move to the $N=8$ case and describe the vacua we obtained.
The algebraic manipulations and the numerical solution of the BPS
differential equations have been performed with the help of
{\it{Mathematica}}.
 In section 6, we make some conclusions.
\section{Three Dimensional Gauged Supergravity}
In this section, we review the basic features of 3 dimensional,
N-extended,  gauged supergravity, following the N-covariant
formulation given in reference \cite{dewit}. We will restrict our
discussion to the case where the scalar manifold is a symmetric
space $G/H$, although for $N< 5$ there are more general
possibilities. Before gauging, the propagating bosonic sector of the
theory is described by  a non-linear sigma model whose target
manifold is $G/H$, where $H$ is a maximal compact subgroup of $G$.
Thus there are scalar fields $\phi^i(x)$, $i=1,\dots,{\rm dim}\,
G/H$, which are coordinates of $G/H$. The subgroup $H$ of $G$
contains the R-symmetry group $SO(N)$.  Gauging proceeds by
introducing  Chern-Simons gauge fields $A_\mu^\mathcal{M}$ in the
adjoint representation of a subgroup $G'$ of the isometry group $G$,
whose embedding in $G$ is specified by a gauge invariant, symmetric
embedding tensor $\Theta_{\mathcal{M}\mathcal{N}}$, with indices
running over the Lie algebra of the gauged subgroup. Supersymmetry
severely restricts the allowed gauged subgroup, and correspondingly
the tensor $\Theta$, as we will see in the following.  For the
reasons explained in the introduction, we will be interested in
non-semisimple gaugings, where the gauged subgroup is a semi-direct
product of a semisimple factor $G_0$ and an abelian factor $T={\bf
R}^{{\rm dim}G_0}$, $G'=G_0\ltimes T$, with  the latter transforming
in the adjoint representation of $G_0$.\\ \indent Let us now
introduce the basic data which allow us to construct the gauged
supergravity theory in the symmetric space case: recall that by
$G/H$ we mean the manifold of right cosets, where $H$ elements
$h(x)$ act by right multiplication on the $G$-valued matrix
$L(\phi^i(x))$. The generators of $G$ decompose into
$\{t^{\mathcal{M}}\}=\{X^{IJ},X^{\alpha},Y^A\}$. $X^{IJ}$ generate
$SO(N)$, and $X^\alpha$ generate a group $H'$ commuting with
$SO(N)$. $Y^A$ are the non-compact generators of $G$. The isometry
group is defined by the left action of $G$ elements on the coset
$G/H$. The geometry of $G/H$  is encoded  in the Lie algebra valued
one-forms $L^{-1} \partial_i L$ and in $L^{-1}t^\mathcal{M}L$,
through the following expansions over Lie algebra generators:
\begin{eqnarray}
L^{-1} \partial_i L&=&
\frac{1}{2}Q^{IJ}_i X^{IJ}+Q^\alpha_i X^{\alpha}+e^A_i Y^A,\nonumber\\
L^{-1}t^\mathcal{M}L&=&\frac{1}{2}\mathcal{V}^{\mathcal{M}IJ}X^{IJ}+\mathcal{V}^\mathcal{M}_{\phantom{a}\alpha}X^\alpha+
\mathcal{V}^\mathcal{M}_{\phantom{a}A}Y^A.
\end{eqnarray}
The $e^A_i$ are vielbeins  which determine the invariant metric
$g_{ij}=e^A_i e^B_j\delta_{AB}$ of $G/H$. The $Q$'s are composite
$H$-connections, and the $\mathcal{V}$'s give the Killing vectors,
$\mathcal{V}^{\mathcal{M}i}=g^{ij}e^{A}_j\mathcal{V}^\mathcal{M}_{\phantom{a}A}$.
Pulling back on space-time and covariantizing with respect to the
gauge action of $G'$ from the right, we define:
\begin{equation}
L^{-1} D_\mu L= \frac{1}{2}Q^{IJ}_\mu X^{IJ}+Q^\alpha_\mu
X^{\alpha}+e^A_\mu Y^A.
\end{equation}
Here $D_\mu L=(\partial_\mu +
\Theta_{\mathcal{M}\mathcal{N}}A_\mu^\mathcal{M}t^\mathcal{N})L$ is
a space-time covariant derivative and it is understood that the
gauge coupling constant is contained in $\Theta$. Thus the full
gauge symmetry of the theory is $L(x)\rightarrow g'(x)L(x)h(x)$,
where $g'\in G'$. The $e^A_\mu$'s give the covariant kinetic term
for scalars,
\begin{equation}
{\mathcal L}_{\textrm{kin}}=\frac{1}{4}\sqrt{g}g^{\mu\nu}e^A_\mu
e^B_\nu\delta_{AB}. \label{kin}
\end{equation}
The Lagrangian for gauge fields is of Chern-Simons type:
\begin{equation}
{\mathcal
L}_{CS}=\frac{1}{4}\epsilon^{\mu\nu\rho}A_\mu^{\mathcal{M}}\Theta_{{\mathcal
M}{\mathcal N}}(\partial_\nu A_{\rho}^{\mathcal N}+
\frac{1}{3}f^{{\mathcal N}{\mathcal P}}_{\mathcal
Q}\Theta_{{\mathcal P}{\mathcal L}} A_\nu^{\mathcal
L}A_\rho^{\mathcal Q}), \label{cs}
\end{equation}
where $f^{{\mathcal N}{\mathcal P}}_{\mathcal Q}$ are the structure
constants of $G'$.\\ \indent As it has been shown in \cite{dewit}
and in more detail in \cite{csym}, in the non-semisimple case where
$G'=G_0\ltimes T$, the Chern-Simons action for $G'$ gauge fields is
equivalent to a Yang-Mills plus Chern-Simons action for gauge fields
transforming under the semisimple part $G_0$. The point is that
gauge invariance implies  that the indices of $\Theta _{{\mathcal
M}{\mathcal N}}$ cannot be both along the $G_0$ direction and this
allows to integrate the gauge fields carrying $G_0$ indices,
producing a Yang-Mills action for gauge fields carrying $T$ indices,
which transform in the usual way under $G_0$. At the same time, one
can use the shift gauge symmetry to remove ${\rm dim}\, G_0$ scalars
from the action.\\ \indent A class of tensors that will play
important role in our analysis are the two-form $SO(N)$ generators,
$f^{IJ}_{ij}$, which originate from the existence of $N-1$ hermitean
almost complex structures $f^{Pi}_{j}$, $P=1,\dots,N-1$, on the
scalar manifold. The existence of the latter is implied by the
existence of $N$ supersymmetries.  They are vector valued one-forms
obeying a Clifford algebra relation and therefore are essentially
$\gamma$-matrices of $SO(N)$. With their commutators one constructs
$SO(N)$ generators $f^{IJ}_{ij}$, which in our case can be expressed
as:
\begin{equation}
f^{IJ}_{ij}=-\Gamma^{IJ}_{AB}e^A_ie^B_j,
\end{equation}
with $\Gamma^{IJ}_{AB}$ properly normalized generators in the spinor
representation of $SO(N)$. Let us now specialize at the $N=4$ and
$N=8$ cases. In the latter case, one proves that the allowed
symmetric spaces are of the form $\frac{SO(8,k)}{SO(8)\times
SO(k)}$, and in fact we will restrict our analysis to $k=8$. For
$N=4$, the scalar manifold can actually be locally the product of
two quaternionic manifolds, and even restricting to the symmetric
space cases, this allows  a finite number of different
possibilities, but we will restrict the analysis to the quaternionic
symmetric space  $\frac {SO(4,4)}{SO(4)\times SO(4)}$.\\ \indent
With the data introduced above, namely the embedding tensor $\Theta$
and the $\mathcal{V}$'s, we define the T-tensors:
\begin{eqnarray}
T^{IJ,KL}&\equiv
&\mathcal{V}^{\mathcal{M}IJ}\Theta_{\mathcal{M}\mathcal{N}}\mathcal{V}^{\mathcal{N}KL},\qquad
T^{IJi}\equiv \mathcal{V}^{\mathcal{M}IJ}\Theta_{\mathcal{M}\mathcal{N}}\mathcal{V}^{\mathcal{N}i},\nonumber\\
T^{ij}&\equiv
&\mathcal{V}^{\mathcal{M}i}\Theta_{\mathcal{M}\mathcal{N}}\mathcal{V}^{\mathcal{N}j},\qquad
T^i_{\phantom{a}\alpha}\equiv \mathcal{V}^\mathcal{M}_{\phantom{a}\alpha}\Theta_{\mathcal{M}\mathcal{N}}\mathcal{V}^{\mathcal{N}i},\nonumber\\
T_{\alpha\beta}&\equiv
&\mathcal{V}^\mathcal{M}_{\phantom{a}\alpha}\Theta_{\mathcal{M}\mathcal{N}}\mathcal{V}^\mathcal{M}_{\phantom{a}\beta},\qquad
T^{IJ}_{\phantom{ab}\alpha}\equiv
\mathcal{V}^{\mathcal{M}IJ}\Theta_{\mathcal{M}\mathcal{N}}\mathcal{V}^\mathcal{N}_{\phantom{a}\alpha}.
\end{eqnarray}
The fundamental consistency constraint on the gauging, implied by
supersymmetry, can be expressed through the following identity:
\begin{equation}
T^{IJ,KL}=T^{[IJ,KL]}-\frac{4}{N-2}\delta^{I[K}T^{L]M,MJ}-\frac{2}{(N-1)(N-2)}\delta^{I[K}\delta^{L]J}T^{MN,MN},
\end{equation}
or equivalently,
\begin{equation}
\mathbb{P}_\boxplus T^{IJ,KL}=0
\end{equation}
which means that the representation $\boxplus$ of $SO(N)$ is
projected out. The scalar potential of the theory can be expressed
in terms of the tensors:
\begin{eqnarray}
A_1^{IJ}&=&-\frac{4}{N-2}T^{IM,JM}+\frac{2}{(N-1)(N-2)}\delta^{IJ}T^{MN,MN},\nonumber\\
A_{2j}^{IJ}&=&\frac{2}{N}T^{IJ}_{\phantom{as}j}+\frac{4}{N(N-2)}f^{M(I
m}_{\phantom{as}j}T^{J)M}_{\phantom{as}m}+\frac{2}{N(N-1)(N-2)}\delta^{IJ}f^{KL\phantom{a}m}_{\phantom{as}j}T^{KL}_{\phantom{as}m}.
\end{eqnarray}
Supersymmetry implies a quadratic identity involving $A_1$ and $A_2$
:
\begin{equation}
2 A_1^{IK}A_1^{KJ}-N A_2^{IKi}
A_{2i}^{JK}=\frac{\delta^{IJ}}{N}(2A_1^{KL}A_1^{KL}-NA_2^{KLi}A_{2i}^{KL}),
\label{qc}
\end{equation}
which offers a non-trivial check on the consistency of the
construction. The scalar potential is given by:
\begin{equation}
V=-\frac{4}{N}\sqrt{g}(A_1^{IJ}A_1^{IJ}-\frac{1}{2}Ng^{ij}A_{2i}^{IJ}A_{2j}^{IJ}).
\end{equation}
Since $\Theta$'s are linear in the gauge couplings, $V$ depends
quadratically on them. The other piece of information we will need
is given by the supersymmetry variations of the matter fermions
$\chi$ and the gravitinos $\psi^I_\mu$. For the former, in order to
use the $SO(N)$ covariant notations, we extend the fermion fields
$\chi^i$ to an overcomplete set $\chi^{iI}$ defined by,\cite{dewit},
\begin{equation}
\chi^{iI}=(\chi^i,f^{Pi}_j\chi^j).
\end{equation}
The Lagrangian and supersymmetry transformation rules can be
expressed in a form that no longer depends explicitly on the almost
complex structures. The fields $\chi^{iI}$ have to satisfy the
projection constraint
\begin{equation}
\chi^{iI}=\mathbb{P}^{Ii}_{Jj}\chi^{jJ}\equiv\frac{1}{N}(\delta^{IJ}\delta^i_j-f^{IJi}_j)\chi^{jJ}.
\end{equation}
Omitting terms which are of  higher order in the fermionic fields,
the supersymmetry transformations which are relevant for us are
given by:
\begin{eqnarray}
\delta\psi^I_\mu
&=&\mathcal{D}_\mu\epsilon^I+A_1^{IJ}\gamma_\mu\epsilon^J,\nonumber\\
\delta\chi^{iI}&=&
\frac{1}{2}(\delta^{IJ}\mathbf{1}-f^{IJ})^i_{\phantom{a}j}{\mathcal{D}{\!\!\!\!/}}\phi^j\epsilon^J
-NA_2^{JIi}\epsilon^J,\label{susyvar}
\end{eqnarray}
where
\begin{eqnarray}
\mathcal{D}_\mu\epsilon^I&=&(\partial_\mu+\frac{1}{2}\omega^a_{\phantom{a}\mu}\gamma_a)\epsilon^I+\partial_\mu\phi^iQ_i^{IJ}\epsilon^J
+\Theta_{\mathcal{M}\mathcal{N}}A_\mu^{\mathcal{M}}\mathcal{V}^{\mathcal{N}IJ}\epsilon^J,\nonumber\\
\mathcal{D}_\mu
\phi^i&=&\partial_\mu\phi^i+A_\mu^{\mathcal{M}}\mathcal{V}^{\mathcal{N}i}\Theta_{\mathcal{M}\mathcal{N}},
\end{eqnarray}
and $\omega^a_\mu$ is the 3-dimensional spin connection constructed
with the dreibein $e^a_\mu$. As shown in \cite{dewit}, assuming a
maximally symmetric space-time (in particular $AdS_3$), the
supersymmetric critical points of the potential are given by the two
equivalent conditions on spinors $\epsilon^I$:
\begin{eqnarray}
A_{2i}^{JI}\epsilon^J&=&0\nonumber\\
\textrm{and}\qquad A_1^{IK}A_1^{KJ}\epsilon^J&=&
-\frac{V_0}{4}\epsilon^I=\frac{1}{N}(A_1^{IJ}A_1^{IJ}-\frac{1}{2}Ng^{ij}A_{2i}^{IJ}A_{2i}^{IJ})\epsilon^I,\label{a1a2condition}
\end{eqnarray}
where $V_0$ is the potential at the critical point. The equivalence
of the two statements follows from the quadratic identity (\ref{qc})
involving $A_1$ and $A_2$. This result says that the preserved
supersymmetries correspond to the eigenvalues of $A_1^{IJ}$ which
equal $\pm\sqrt{\frac{-V_0}{4}}$, since in our normalization $-V_0=R^{-2}$,
where $R$ is the radius of $AdS_3$.
More in detail, let us  choose $AdS_3$ coordinates $r, x_0,x_1$,
and metric $ds^2=dr^2+
e^{2r/R}(-dx_0^2+dx_1^2)$.  From the previous remarks, it follows
that for each  eigenvector $v ^I_{\pm}$ of
$A_1^{IJ}$, with eigenvalue  $\pm\sqrt{\frac{-V_0}{4}}$, if we form the
spinor $\epsilon^I_{\pm}=\epsilon_{\pm}\otimes v^I_{\pm}$, then the BPS condition
for the gravitino variation (\ref{susyvar}) becomes identical to the Killing spinor equation
for $\epsilon_{\pm}$ on $AdS_3$ i.e.  $\mathcal{D}_\mu \epsilon_{\pm}=\pm \frac{1}{2R} \gamma_\mu\epsilon_{\pm}$.
Using the explicit expression for the spin connection
for the above metric, one can see that one solution to this equation is an $x_0,x_1$-independent
spinor obeying $\gamma^r\epsilon_{\pm}=
\pm \epsilon_{\pm}$, where $\gamma^r$ is the flat gamma matrix. This
corresponds to a left (right) Poincare' supersymmetry in the boundary CFT.
The other solution gives rise to the superconformal charge in the boundary CFT, has a non-trivial
$x_0,x_1$ dependence and is constructed with a constant spinor obeying the
opposite $\gamma^r$ projection condition.
\\ \indent Therefore, it is convenient to classify the critical points by
presenting their preserved supersymmetries in the form of
$(N_+,N_-)$ corresponding to the $N_+$ and $N_-$ positive and
negative eigenvalues of $A_1^{IJ}$ whose modulus equals
$\sqrt{\frac{-V_0}{4}}$. These coincide with the number of
left-(right-) moving Poincare' supersymmetries of the dual $SCFT_2$.
Of course the total number of supersymmetries is doubled by the
inclusion of the superconformal ones.
\\ \indent
To summarize, the procedure of
finding supersymmetric vacua is the following. From
\eqref{a1a2condition}, we look for the Killing spinors $\epsilon^I$
which are annihilated by some of the $A_{2i}^{JI}$. At the same
time, $\epsilon^I$ must also be the eigenvector of $A_1^{IJ}$.
Clearly, maximal supersymmetric vacua are annihilated by all of the
components of $A_{2i}^{JI}$, and $\epsilon^I$ is an eigenvector of
$A_1^{IJ}$ for all directions $I$. The $\epsilon^I$ characterizing
partially supersymmetric vacua will be an eigenvector of $A_1^{IJ}$
for certain directions labeled by some values of $I$, and will be
annihilated only by the $A_{2i}^{JI}$ in the corresponding
directions. We also find many supersymmetric vacua with $V_0=0$, and
there might be non-supersymmetric $AdS_3$ vacua as well. However, in
this work, we will not discuss them.
\section{Vacua of the $N=4$ Theory }
The target space in our case is the product of two quaternionic
manifolds, that we take to be $SO(4,4)/SO(4)\times SO(4)$. A
convenient (redundant) parametrization of cosets is given  by the
following SO(4,4) group element
\begin{equation}
L_i=\frac{1}{2}\left(
    \begin{array}{cc}
      X_i+e^t_i & Y_i+e^t_i \\
      -X_i+e^t_i & e^t_i-Y_i \\
    \end{array}
  \right),
\end{equation}
where $i=1,2$ refers to the two spaces. $e_i$ is a $4\times 4$
matrix in GL(4,{\bf R}), $X_i=E_i+B_ie_i^t$, $Y_i=-E_i+B_ie_i^t$.
$B_i$ is an antisymmetric $4\times 4$ matrix, and $E_i=e_i^{-1}$.
The inverse of $L_i$ is
\begin{equation}
L^{-1}_i=\frac{1}{2}\left(
         \begin{array}{cc}
           X^t_i+e_i & X^t_i-e_i \\
           -Y^t_i-e_i & e_i-Y_i^t \\
         \end{array}
       \right).
\end{equation}
One can eliminate 6 of the 22 parameters in $L$ by using the right
action of the diagonal $SO(4)$ action, for example by bringing $e_i$
into an upper triangular form. The following Lie algebra elements,
\begin{equation}
t^\mathcal{A}=\left(
                \begin{array}{cc}
                  a & 0 \\
                  0 & a \\
                \end{array}
              \right)\qquad
 t^\mathcal{B}=\left(
                 \begin{array}{cc}
                   b & b \\
                   -b & -b \\
                 \end{array}
               \right)\label{gauging}
 \end{equation}
where all entries are $4\times 4$ antisymmetric blocks, together
with an identical copy for the second space, will be gauged. In
other words, the semisimple part of the gauge group will be the
diagonal $SO(4)_D$ in the $(SO(4))^4$ of the product
$(\frac{SO(4,4)}{SO(4)\times SO(4)})^2$, corresponding to generators
$t^\mathcal{A}$. On the other hand, the nilpotent generators,
$t^\mathcal{B}$, generate diagonal shift symmetries
$B_{1,2}\rightarrow B_{1,2}+2b$. Also, it is clear that the
$\mathcal{B}$-generators transform in the adjoint representation
with respect to the diagonal $SO(4)$. For $a$ and $b$, we can take a
basis of antisymmetric matrices given by
$J^{IJ}=\varepsilon^{IJ}-\varepsilon^{JI}$, with
$(\varepsilon^{IJ})_{KL}=\delta_{IK}\delta_{JL}$. Similarly, we can
use the following basis for the 16 non-compact generators of
$SO(4,4)$:
\begin{equation}
Y^{ab}=\left(
         \begin{array}{cc}
           0 & \varepsilon^{ab} \\
           (\varepsilon^t)^{ab} & 0 \\
         \end{array}
       \right).
\end{equation}
Since in the present case both the R symmetry group and the gauge
group are $SO(4)$, it is convenient to split the corresponding Lie
algebras generators into self-dual and anti-self-dual components
$J_+$ and $J_-$ respectively:
\begin{equation}
J_+^{IJ}=J^{IJ}+\frac{1}{2}\epsilon^{IJKL}J^{KL} \qquad
\textrm{and}\qquad J_-^{IJ}=J^{IJ}-\frac{1}{2}\epsilon^{IJKL}J^{KL}
\end{equation}
which are $SU(2)_+$ and $SU(2)_-$ generators in the
$SO(4)=SU(2)_+\oplus SU(2)_-$ Lie algebra decomposition. We will
adopt this decomposition both for $\mathcal A$- and $\mathcal
B$-type generators. Correspondingly, the two-forms tensors $f^{IJ}$
introduced in the previous section have, say, self-dual components
on the first quaternionic space and anti-self-dual components on the
second. In our formalism and in a flat basis, they can be expressed
as:
\begin{equation}
{f^{IJ}_{\pm}}_{ab,cd}=\textrm{Tr}((\varepsilon^t)^{ab}J_{\pm}^{IJ}\varepsilon^{cd}).
\end{equation}
At this stage, we can proceed to construct the supergravity theory
with the gauging of $SO(4)\ltimes {\bf R}^6$ and in particular,
verify its consistency, along the lines reviewed in the previous
section. As explained there, the main ingredients are given by the
tensors $A_1$ and $A_2$, which determine the scalar potential and
the supersymmetry variations of the fermionic fields. They are
constructed through the $T$-tensors, which in turn are obtained by
uplifting the embedding tensor $\Theta_{\mathcal{M}\mathcal{N}}$
into $G$ by using ${\mathcal V}^{\mathcal{M}}_{\mathcal{P}}$, with
$\mathcal{P}$ running over the generators of $G$ corresponding to
the R-symmetries $\mathcal{P}=IJ$, and the non-compact coset
directions $\mathcal{P}=ab$ in the first and second space. We give
in the Appendix \ref{formulae} expressions for the relevant
components of $\mathcal V$.
\\ \indent Gauge invariance restricts the $\Theta$ tensors to have components,
$\Theta_{\mathcal{A}\mathcal{B}}$ and
$\Theta_{\mathcal{B}\mathcal{B}}$, which are proportional to the
$SO(4)$ Killing form, schematically
$\delta_{\mathcal{A}\mathcal{B}}$ and
$\delta_{\mathcal{B}\mathcal{B}}$, respectively. The proportionality
constants are gauge couplings, and, of course, we should specify
here to which of the four $SU(2)$'s the $\mathcal{A}$, $\mathcal{B}$
indices belong. Therefore, a priori we expect  four  couplings
$g_{1s}, g_{1a}, g_{2s},$ and $g_{2a}$. The $a$ and $s$ labels
indicate the self-dual and anti-self-dual $SU(2)$, respectively, and
1 refers to the $\mathcal{A}\mathcal{B}$ couplings whereas 2 refers
to the $\mathcal{B}\mathcal{B}$ ones. \\ \indent With this notation
and with the meaning of $\mathcal V$ indices explained in the
Appendix \ref{formulae}, the $T$-tensors turn out to be:
\begin{eqnarray}
T^{LJ,MK}&=&g_{1s}(\mathcal{V}_{\textrm{+a}}^{LJ,PQ}\mathcal{V}_{\textrm{+b}}^{MK,PQ}+\mathcal{V}_{\textrm{+b}}^{LJ,PQ}\mathcal{V}_{\textrm{+a}}^{MK,PQ})+
g_{1a}(\mathcal{V}_{-\textrm{a}}^{LJ,PQ}\mathcal{V}_{-\textrm{b}}^{MK,PQ}\nonumber\\&
& +
\mathcal{V}_{-\textrm{b}}^{LJ,PQ}\mathcal{V}_{-\textrm{a}}^{MK,PQ})
+g_{2s}\mathcal{V}_{\textrm{+b}}^{LJ,PQ}\mathcal{V}_{\textrm{+b}}^{MK,PQ}+g_{2a}\mathcal{V}_{-\textrm{b}}^{LJ,PQ}\mathcal{V}_{-\textrm{b}}^{MK,PQ},\nonumber\\
{T_1}^{LJ}_{ab}&=&g_{1s}(\mathcal{V}_{\textrm{+a}}^{LJ,PQ}{\mathcal{V}_{\textrm{+1b}}}^{PQ}_{ab}+\mathcal{V}_{\textrm{+b}}^{LJ,PQ}{\mathcal{V}_{\textrm{+1a}}}
^{PQ}_{ab}) +g_{1a}(\mathcal{V}_{-\textrm{a}}^{LJ,PQ}
{\mathcal{V}_{-\textrm{1b}}}^{PQ}_{ab}\nonumber\\ & &+
\mathcal{V}_{-\textrm{b}}^{LJ,PQ}{\mathcal{V}_{-\textrm{1a}}}^{PQ}_{ab})
+g_{2s}\mathcal{V}_{\textrm{+b}}^{LJ,PQ}
{\mathcal{V}_{\textrm{+1b}}}^{PQ}_{ab}+g_{2a}\mathcal{V}_{-\textrm{b}}^{LJ,PQ}{\mathcal{V}_{-\textrm{1b}}}^{PQ}_{ab},
\nonumber \\
{T_2}^{LJ}_{ab}&=&g_{1s}(\mathcal{V}_{\textrm{+a}}^{LJ,PQ}{\mathcal{V}_{\textrm{+2b}}}^{PQ}_{ab}+\mathcal{V}_{\textrm{+b}}^{LJ,PQ}{\mathcal{V}_{\textrm{+2a}}}
^{PQ}_{ab}) +g_{1a}(\mathcal{V}_{-\textrm{a}}^{LJ,PQ}
{\mathcal{V}_{-\textrm{2b}}}^{PQ}_{ab}\nonumber\\&
&+\mathcal{V}_{-\textrm{b}}^{LJ,PQ}{\mathcal{V}_{-\textrm{2a}}}^{PQ}_{ab})
+g_{2s}\mathcal{V}_{\textrm{+b}}^{LJ,PQ}
{\mathcal{V}_{\textrm{+2b}}}^{PQ}_{ab}+g_{2a}\mathcal{V}_{-\textrm{b}}^{LJ,PQ}{\mathcal{V}_{-\textrm{2b}}}^{PQ}_{ab}.
\label{T}
\end{eqnarray}
\indent It turns out that the consistency requirement on
$T^{IJ,KL}$, discussed in the previous section, requires
$g_{2a}=-g_{2s}$. Moreover, we find it is convenient for the
subsequent analysis to redefine the couplings from $g_{1s}$,
$g_{1a}$ to $g_n$, $g_p$ as follows:
\begin{equation}
g_{1s}=g_p+g_n\qquad \textrm{and}\qquad g_{1a}=g_p-g_n.
\end{equation}
\indent Now, we study various vacua of this theory. We begin by
choosing an ansatz for the coset $L$. We have two spaces. We set
$B_1=B_2=0$ and choose diagonal $e_i$'s:
\begin{equation}
e_1=\left(
      \begin{array}{cccc}
        a_1 & 0 & 0 & 0 \\
        0 & a_2 & 0 & 0 \\
        0 & 0 & a_3 & 0 \\
        0 & 0 & 0 & a_4 \\
      \end{array}
    \right) \qquad \textrm{and}\qquad
e_2=\left(
      \begin{array}{cccc}
        b_1 & 0 & 0 & 0 \\
        0 & b_2 & 0 & 0 \\
        0 & 0 & b_3 & 0 \\
        0 & 0 & 0 & b_4 \\
      \end{array}
    \right).
\end{equation}
Notice that the shift gauge symmetry would allow us to set one of
the two $B$'s to zero and the  left $SO(4)$ gauge symmetry can be
used to diagonalize one of the two $e$'s, so the ansatz above is
indeed a truncation of the full twenty-dimensional moduli space. We
have checked the consistency of this truncation explicitly. That is,
 we have verified that the remaining fields appear
at least quadratically in the action, and therefore setting them to
zero solves their equations of motion. We then proceed to analyze
the BPS conditions $\delta \psi^I_\mu=0$ and $\delta\chi^{iI}=0$
using (\ref{susyvar}), within this eight-dimensional subspace.
\\ \indent We give below the vacuum expectation values of $e_1$, $e_2$, the
$A_1^{IJ}$ eigenvalue ($A_1$) satisfying $|A_1|^2=- V_0/4$ and the
corresponding preserved supersymmetries $(N_+,N_-)$ for the $AdS_3$
vacuum solutions that are relevant to the flow solutions we will
show in the next section. Other vacua are shown in Appendix
\ref{vacua}.
\subsection{(3,1) vacua}
\begin{itemize}
  \item I.
  \begin{eqnarray}
  e_1&=&\sqrt{\frac{-2(g_n+g_p)}{g_{2s}}}\mathbb{I}_{4\times 4}\nonumber\\
  e_2&=&\sqrt{\frac{-2(g_n+g_p)}{g_{2s}}}(-1,1
  ,1,1)\nonumber\\
  A_1&=&\frac{32(g_n+g_p)^2}{g_{2s}}\qquad \textrm{and}\qquad
  V_0=\frac{-4096(g_n+g_p)^4}{g_{2s}^2}.
  \end{eqnarray}
  \item II.
  \begin{eqnarray}
  e_1&=&\sqrt{\frac{2(g_p-g_n)}{g_{2s}}}(1,-1
  ,-1,-1)\nonumber\\
  e_2&=&-\sqrt{\frac{2(g_p-g_n)}{g_{2s}}}\mathbb{I}_{4\times 4}\nonumber\\
  A_1&=&\frac{-32(g_n-g_p)^2}{g_{2s}}\qquad \textrm{and}\qquad
  V_0=\frac{-4096(g_n-g_p)^4}{g_{2s}^2}.
  \end{eqnarray}
  \item III.
  \begin{eqnarray}
  e_1&=&\sqrt{\frac{g_n(g_p^2-g_n^2)}{g_{2s}g_n^2}}\Big(\frac{g_n}{g_p},-1
  ,-1,-1\Big)\nonumber\\
  e_2&=&-\sqrt{\frac{g_n(g_p^2-g_n^2)}{g_{2s}g_n^2}}\Big(\frac{g_n}{g_p},1
  ,1,1\Big)\nonumber\\
  A_1&=&\frac{-8(g_n^2-g_p^2)^2}{g_{2s}g_ng_p}\qquad \textrm{and}\qquad
  V_0=\frac{-256(g_n^2-g_p^2)^4}{g_{2s}^2g_n^2g_p^2}.
  \end{eqnarray}
\end{itemize}
\subsection{(2,0) vacua}
\begin{itemize}
\item IV.
\begin{equation}
  e_1=(-a_1, a_1, a_2, a_2)\qquad e_2=(b_1, b_1, b_2, b_2)
 \end{equation}
\begin{eqnarray}
   a_1&=&2\sqrt{\frac{g_p^2-g_n^2}{g_{2s}(g_p-g_n+\sqrt{5g_n^2+2g_pg_n+g_p^2})}}\nonumber\\
a_2&=&2\sqrt{\frac{g_p^2-g_n^2}{g_{2s}(g_n-g_p+\sqrt{5g_p^2+2g_pg_n+g_n^2})}}\nonumber\\
b_1&=&2\sqrt{\frac{g_p^2-g_n^2}{g_{2s}(3g_n+g_p+\sqrt{5g_n^2+2g_pg_n+g_p^2})}}\nonumber\\
b_2&=&2\sqrt{\frac{g_p^2-g_n^2}{g_{2s}(\sqrt{g_n^2+2g_ng_p+5g_p^2}+g_n+3g_p)}}
\end{eqnarray}
  \begin{eqnarray} A_1&=&\frac{-32(g_n-g_p)^2}{g_{2s}}\qquad\textrm{and}\qquad
V_0=-\frac{4096(g_n-g_p)^4}{g_{2s}^2}.
\end{eqnarray}
 \item V.
  \begin{eqnarray}
  e_1&=&(a_1, a_2, a_3, a_3)\qquad e_2=(b_1, b_1, b_2, b_2)\nonumber\\
   \end{eqnarray}
  \begin{eqnarray}
  a_1&=&-\frac{1+t}{1-t+\sqrt{1+t^2}}\sqrt{\frac{2g_p(1-t+\sqrt{1+t^2})}{g_{2s}t(1+t)\sqrt{1+t^2}}}\times \nonumber\\
  & &\sqrt{(t-1) \left\{t^3-t^2+t-1+(t-t^2-1)\sqrt{1+t^2}\right\}}\nonumber\\
  a_2&=&\sqrt{\frac{2tg_p(t-1)^2(1+t)\sqrt{1+t^2}}{g_{2s}(1-t+\sqrt{1+t^2})}}\times\nonumber\\
 & & \frac{1}{\sqrt{(t-1-t^2)(t-1)\sqrt{1+t^2}-t^2+(1-t+t^2)^2}}
  \nonumber\\
  a_3&=&\sqrt{\frac{2g_p(1-t^2)}{g_{2s}(t-1+\sqrt{1+t^2})}}\nonumber\\
b_1&=&\sqrt{\frac{2g_p(1-t^2)}{g_{2s}(1+t+\sqrt{1+t^2})}}\nonumber\\
   b_2&=&\sqrt{\frac{2g_p(1-t^2)}{g_{2s}(1+t+\sqrt{1+t^2})}}
   \nonumber\\ A_1&=&\frac{-8(g_n^2-g_p^2)^2}{g_{2s}g_ng_p}\qquad\textrm{and}\qquad
V_0=-\frac{256(g_n^2-g_p^2)^4}{(g_{2s}g_ng_p)^2},
\end{eqnarray}
where we have introduced $t=\frac{g_n}{g_p}$.
\end{itemize}
\indent Out of all vacua, there are only three possibilities in
connecting two vacua. That means we will have only three RG flows in
the dual field theories. All these three flows are the flows between
I and III, II and III, and between IV and V. The last flow is the
only possible flow among V and other (2,0) points. This is because
we cannot find any values of $g_n$, $g_p$ and $g_{2s}$ so that both
$e_1$ and $e_2$ of the two end points of the flow are real apart
from the IV and V pair. There are three possibilities in order to
make IV and V real at the same time. These are given by
\begin{eqnarray}
& &t<-1, g_p<0, g_{2s}<0 \nonumber\\
\textrm{or}\qquad & & t<1, g_p>0, g_{2s}>0\nonumber\\
\textrm{or}\qquad & & t>1, g_p>0, g_{2s}<0.
\end{eqnarray}
For definiteness, we choose the last range and further choose $t=2$,
$g_p=1$ and $g_{2s}=-1$ in our numerical solution. For all the
critical points given above, we have checked that there exist at
least one possible set of $g_p$, $g_n$ and $g_{2s}$ such that all
the square roots in any critical points are real, although any two
different critical points may not be made real with the same values
of $g_p$, $g_n$ and $g_{2s}$.
\\ \indent There might be more possibilities apart from these three
flows. However, we could not find any interpolating solutions both
analytically and numerically apart from those three mentioned above.
Remarkably, we find only the flows between critical points which
have the same supersymmetries. In the next section, we will give
these solutions explicitly.
\section{Supersymmetric Flow Solutions}
In this section, we study flows between some pairs of $AdS$ vacua
found in the previous section. We assume the standard form for the
3D metric:
\begin{equation}
ds^2=e^{2A(r)}(-dt^2+dx^2)+dr^2.
\end{equation}
This becomes the $AdS_3$ metric for $A(r)=r/R$, where  $R$ is the
$AdS_3$ radius. This is related to the vacuum energy $V_0$ as
$R^2=-1/V_0$, since in our normalization Einstein's equations read
$R_{\mu\nu}=-2 V_0 g_{\mu\nu}$. Also, we recall that the eigenvalue
$A_1$ introduced in the previous section satisfies $4 A_1^2=-V_0$.
We will look for solutions of the BPS equations interpolating
between $AdS$ vacua from the UV region ($r\rightarrow +\infty$) to
the IR region ($r\rightarrow -\infty$), where the scalar fields
reach the vev's determined in the previous section. The central
charge of the CFT's at an $AdS_3$ vacuum is proportional to $R$, and
therefore proportional to $1/A'(r)$. In fact, the latter quantity
can be used to define, up to a positive proportionality constant, a
$C$-function, $C(r)$, on the full flow interpolating between the UV
and IR fixed points and can be proved to be monotonic, $A''(r)\leq
0$ \cite{fgpw}. This nicely agrees with the c-theorem in conformal
field theories. The result in \cite{fgpw} depends on the validity of
the weaker energy condition, which is met in  all the flows
involving only scalars and the metric. This is the case for our
flows as we will see below. Notice that, since $A(r)$ is related to
$A_1$ through a first order differential equation given by the
gravitino variation (\ref{susyvar}), this also implies that $A_1$
should not change sign along the flow because this would imply an
unphysical infinity for $C(r)$ at some value of $r$. Examples of RG
flows in 3D gauged supergravity have been studied in \cite{bs,
deger}.
\subsection{The Flow Between $(3,1)$ Vacua}
In this subsection, we study a supersymmetric flow between two of
the $AdS_3$ vacua with the same, $(3,1)$, amount of supersymmetries
but with different cosmological constants, found in the previous
section.\\ \indent We start by giving an ansatz for the scalars with
non-trivial $r$-dependence,
\begin{equation}
e_1=\left(
      \begin{array}{cccc}
        b(r) & 0 & 0 & 0 \\
        0 & a(r) & 0 & 0 \\
        0 & 0 & a(r) & 0 \\
        0 & 0 & 0 & a(r) \\
      \end{array}
    \right),
\qquad e_2=\left(
      \begin{array}{cccc}
        -b(r) & 0 & 0 & 0 \\
        0 & a(r) & 0 & 0 \\
        0 & 0 & a(r) & 0 \\
        0 & 0 & 0 & a(r) \\
      \end{array}
    \right).
\end{equation}
Since now we are going to allow the scalars to have $r$ dependence,
we need to worry about possible contributions of the intrinsic
connection $Q_\mu^{IJ}$ and the gauge fields $A_\mu^{\mathcal M}$ to
the BPS equations (\ref{susyvar}). In addition, of course, the
Yang-Mills equations of motion may be non-trivial. Indeed,
$r$-dependent scalars may a priori source the gauge fields in case
they give rise to a non-trivial gauge current $J_\mu^{\mathcal M}$.
 From the kinetic term (\ref{kin}), we have
\begin{eqnarray}
{\mathcal L}_{\textrm{kin}}&=&\frac{1}{2}\sqrt{g}[
\textrm{Tr}(L^{-1}\partial_\mu LL^{-1}\partial^\mu
L)+2\Theta_{\mathcal{M}\mathcal{N}}A^{\mathcal{M}\mu}\textrm{Tr}(L^{-1}t^{\mathcal{N}}\partial_\mu
L) \nonumber\\&
&+\Theta_{\mathcal{M}\mathcal{N}}\Theta_{\mathcal{K}\mathcal{L}}A^{\mathcal{M}\mu}A_{\mu}^{\mathcal{K}}
\textrm{Tr}(L^{-1}t^{\mathcal{N}}t^{\mathcal{L}}L)].\label{kin1}
\end{eqnarray}
From (\ref{kin1}), we see that the gauge fields couple to the scalar
fields via a current
\begin{equation}
J_\mu^{\mathcal{N}}=\sqrt{g}\textrm{Tr}(L^{-1}t^{\mathcal{N}}\partial_\mu
L).
\end{equation}
For diagonal $e_1$ and $e_2$, the current is zero, so we can
consistently satisfy the equation of motion for the gauge fields by
setting $A_\mu^{\mathcal{M}}=0$. As promised, our flows involve only
scalars and the metric. So, the holographically proved c-theorem
mentioned before is guaranteed in our flow ansatz. Furthermore, all
of the composite connections $Q$'s are also zero in this diagonal
ansatz. The BPS equations can be obtained by using (\ref{susyvar}).
The $\delta \chi^{iI}=0$ conditions give
\begin{eqnarray}
\frac{db}{dr}&=&24g_n ab^2+16g_pb^3-8a^3(g_n-g_{2s}b^2)\label{aeq}\\
\frac{da}{dr}&=&16g_pa^3+8g_na^2b+\frac{8a^4(g_n+g_{2s}b^2)}{b}.\label{beq}
\end{eqnarray}
This ansatz preserves (3,1) supersymmetry, so we have (3,1)
supersymmetry throughout the flow. We proceed by taking one of the
scalars as an independent variable. Changing the variables to
$b(r)=z$ and $a(r)=a(z)$, we can write (\ref{aeq}) and (\ref{beq})
as a single equation
\begin{equation}
\frac{da}{dz}=\frac{a^2(g_nz^2+2g_pza+(g_n+g_{2s}z^2)a^2)}{2g_pz^4+3g_nz^3a+(g_{2s}z^3-g_nz)a^3}.\label{eq}
\end{equation}
We solve this by writing $a(z)=z f(z)$. Then, (\ref{eq}) becomes
\begin{equation}
z\frac{df}{dz}=-\frac{2f(g_p+g_nf)(f^2-1)}{(g_n-g_{2s}z^2)f^3-2g_p-3g_nf}.
\end{equation}
This equation can be solved for $z$ as a function of $f$. We find
\begin{equation}
z=\pm\sqrt{\frac{g_n(f^2-1)}{g_{2s}f^2+(g_n^2f^3+g_ng_pf^2)c_1}}.
\end{equation}
We then obtain
\begin{eqnarray}
b&=&\pm\sqrt{\frac{g_n(f^2-1)}{g_{2s}f^2+(g_n^2f^3+g_ng_pf^2)c_1}},\\
a&=&f b,
\end{eqnarray}
and (\ref{aeq}) and (\ref{beq}) lead to the same equation for $f$
\begin{equation}
\frac{df}{dr}=\frac{16g_n(g_p+g_nf)(f^2-1)^2}{f(g_{2s}+(g_ng_p+g_n^2f)c_1)}.
\end{equation}
We can solve for $r$ in term of $f$ and find
\begin{eqnarray}
r&=&c_2+\frac{1}{64g_n}\bigg[\frac{2(-fg_{2s}g_n+g_{2s}g_p+g_n(g_p^2-g_n^2)c_1)}{(f^2-1)(g_n^2-g_p^2)}-\frac{g_{2s}g_n\ln(1-f)}{(g_n+g_p)^2}
\nonumber\\&
&+\frac{g_{2s}g_n\ln(1+f)}{(g_n-g_p)^2}-\frac{4g_{2s}g_n^2g_p\ln(fg_n+g_p)}{(g_n^2-g_p^2)^2}\bigg].
\end{eqnarray}
The constant $c_2$ is irrelevant and can be set to zero by shifting
the coordinate $r$. So, from now on, we will use $c_2=0$ and choose
a definite sign, $+$ sign, for $z$. \\ \indent We now move to the
gravitino variation $\delta \psi^I_\mu$. The BPS condition gives an
equation for the warp factor $A(r)$:
\begin{eqnarray}
\frac{dA}{dr}&=&
-\frac{1}{f^2(g_{2s}+(g_ng_p-g_n^2f)c_1)^2}[16g_n(f^2-1)(3f^2(c_1g_n(g_n^2+g_p^2)+g_{2s}g_p)\nonumber\\
& &-2g_nf^3(2c_1g_ng_p+g_{2s})
-2g_nf(2c_1g_ng_p+g_{2s})+c_1g_n^3f^4\nonumber\\
& &+g_p(c_1g_ng_p+g_{2s}))].
\end{eqnarray}
Changing the variable from $r$ to $f$, we find
\begin{equation}
\frac{dA}{df}=\frac{1}{fg_n+g_p}\bigg[\frac{g_p+f(3fg_p+g_n(3+f^2))}{f(f^2-1)}-\frac{g_{2s}g_n}{g_{2s}+g_n(fg_n
+g_p)c_1}\bigg].
\end{equation}
This can be solved and give
\begin{equation}
A=c_3+\ln f-2\ln(1-f^2)+\ln(g_p+fg_n)+\ln(g_{2s}+g_n(g_p+g_nf)c_1).
\end{equation}
The constant $c_3$ can be set to zero by rescaling coordinates $x^0$
and $x^1$. We require that $A_1$ must not change sign along the
flow, so these are the only two possible flows namely the flow
between I and III critical points and between II and III points. We
choose the value of $c_1$ in such a way that the solution goes to
one critical point at one end and to the other critical point at the
other end. In order to identify the UV point with $r=\infty$ and the
IR point with $r=-\infty$, we choose $g_{2s}<0$ in the followings.
\\ \indent In the flow between I and III  critical
points, we chose $c_1=-\frac{g_{2s}}{g_n(g_n+g_p)}$, $g_ng_p<0$ and
obtain
\begin{eqnarray}
b&=&\sqrt{-\frac{(g_n+g_p)(1+f)}{g_{2s}f^2}}\nonumber\\
a&=&\sqrt{-\frac{(g_n+g_p)(1+f)}{g_{2s}}}\nonumber\\
r&=&\frac{1}{64}\bigg[-\frac{2g_{2s}}{(1+f)(g_n^2-g_p^2)}-\frac{g_{2s}\ln(1-f)}{(g_n+g_p)^2}
\nonumber\\&
&+\frac{g_{2s}\ln(1+f)}{(g_n-g_p)^2}-\frac{4g_{2s}g_ng_p\ln(fg_n+g_p)}{(g_n^2-g_p^2)^2}\bigg]\nonumber\\
A&=&\ln f-\ln(1-f)-2\ln(1+f)+\ln(g_p+fg_n)
\end{eqnarray}
where we have absorbed all the constants in $c_3$ for the last
equation. We see that $A\rightarrow \infty$ at $f=1$ and
$A\rightarrow -\infty$ at $f=-\frac{g_p}{g_n}$. In the dual CFT, the
I point corresponds to the UV fixed point while the III point
corresponds to the IR point. The ratio of the central charges is
given by
\begin{equation}
\frac{c_{UV}}{c_{IR}}=-\frac{(g_n-g_p)^2}{4g_ng_p}.
\end{equation}
It is easy to show that this is always greater than 1 as it should.
\\ \indent The flow between II and III are given by
$c_1=\frac{g_{2s}}{g_n(g_n-g_p)}$, and $g_ng_p>0$. We find that
\begin{eqnarray}
b&=&\sqrt{\frac{(g_n-g_p)(f-1)}{g_{2s}f^2}}\nonumber\\
a&=&\sqrt{\frac{(g_n-g_p)(f-1)}{g_{2s}}}\nonumber\\
r&=&\frac{1}{64}\bigg[\frac{2g_{2s}}{(1-f)(g_n^2-g_p^2)}-\frac{g_{2s}\ln(1-f)}{(g_n+g_p)^2}
\nonumber\\&
&+\frac{g_{2s}\ln(1+f)}{(g_n-g_p)^2}-\frac{4g_{2s}g_ng_p\ln(fg_n+g_p)}{(g_n^2-g_p^2)^2}\bigg]\nonumber\\
A&=&\ln f-2\ln(1-f)-\ln(1+f)+\ln(g_p+fg_n).
\end{eqnarray}
In this case, we see that $A\rightarrow \infty$ at $f=-1$ and
$A\rightarrow -\infty$ at $f=-\frac{g_p}{g_n}$. In the dual CFT, the
II point corresponds to the UV fixed point while the III point
corresponds to the IR point. The ratio of the central charges is
given by
\begin{equation}
\frac{c_{UV}}{c_{IR}}=\frac{(g_n+g_p)^2}{4g_ng_p}.
\end{equation}
Again,this agrees with the c-theorem.
\\ \indent We next compute the scalar mass spectrum for the eight scalars. We parametrize the eight scalars as follow:
\begin{eqnarray}
a_1(r)&=&a_{10}e^{s_1(r)}\qquad a_2(r)=a_{20}e^{s_2(r)} \nonumber\\
a_3(r)&=&a_{30}e^{s_3(r)}\qquad a_4(r)=a_{40}e^{s_4(r)} \nonumber\\
b_1(r)&=&a_{50}e^{s_5(r)}\qquad b_2(r)=a_{60}e^{s_6(r)} \nonumber\\
b_3(r)&=&a_{70}e^{s_7(r)}\qquad b_4(r)=a_{80}e^{s_8(r)}
\end{eqnarray}
where all the $s_i$, $i=1,\ldots 8$ are canonically normalized
scalars. From the scalar mass matrix $M^2$, we can find the
conformal dimensions ($\Delta$) of the operators in the dual CFT by
using the relation
\begin{equation}
\Delta (\Delta-2)=m^2R^2.
\end{equation}
We find the following mass matrices.
\begin{itemize}
  \item $f=1$:
  \begin{equation}
  M^2=\frac{2048 (g_n+g_p)^4}{g_{2s}^2}\left(
\begin{array}{cccccccc}
 0 & 1 & 1 & 1 & 0 & 1 & 1 & 1 \\
 1 & 0 & 1 & 1 & 1 & 0 & 1 & 1 \\
 1 & 1 & 0 & 1 & 1 & 1 & 0 & 1 \\
 1 & 1 & 1 & 0 & 1 & 1 & 1 & 0 \\
 0 & 1 & 1 & 1 & 0 & 1 & 1 & 1 \\
 1 & 0 & 1 & 1 & 1 & 0 & 1 & 1 \\
 1 & 1 & 0 & 1 & 1 & 1 & 0 & 1 \\
 1 & 1 & 1 & 0 & 1 & 1 & 1 & 0
\end{array}
\right).
  \end{equation}
  The eigenvalues of $M^2R^2$ are (3, -1, -1, -1, 0, 0, 0, 0) corresponding to $\Delta=(3, 1, 2)$.
  All the eight eigenvectors are given by
  \begin{eqnarray}
  v_1&=&(1, 1, 1, 1, 1, 1, 1, 1)\qquad v_2=(-1, 0, 0, 1, -1, 0, 0, 1)\nonumber\\
  v_3&=&(-1, 0, 1, 0, -1, 0, 1, 0)\qquad v_4=(-1, 1, 0, 0, -1, 1, 0, 0)\nonumber\\
  v_5&=&(0, 0, 0, -1, 0, 0, 0, 1)\qquad v_6=(0, 0, -1, 0, 0, 0, 1, 0)\nonumber\\
  v_7&=&(0, -1, 0, 0, 0, 1, 0, 0)\qquad v_8=(-1, 0, 0, 0, 1, 0, 0, 0).
  \end{eqnarray}
  Our flow corresponds to the combination $v_2+v_3+v_4$ which has eigenvalue -1, $\Delta=1$. This is consistent with the fact that the flow
  is driven by a relevant operator.
  \item $f=-1$:
  \begin{equation}
  M^2=\frac{2048 (g_n-g_p)^4}{g_{2s}^2}\left(
\begin{array}{cccccccc}
 0 & 1 & 1 & 1 & 0 & 1 & 1 & 1 \\
 1 & 0 & 1 & 1 & 1 & 0 & 1 & 1 \\
 1 & 1 & 0 & 1 & 1 & 1 & 0 & 1 \\
 1 & 1 & 1 & 0 & 1 & 1 & 1 & 0 \\
 0 & 1 & 1 & 1 & 0 & 1 & 1 & 1 \\
 1 & 0 & 1 & 1 & 1 & 0 & 1 & 1 \\
 1 & 1 & 0 & 1 & 1 & 1 & 0 & 1 \\
 1 & 1 & 1 & 0 & 1 & 1 & 1 & 0
\end{array}
\right).
  \end{equation}
   The eigenvalues of $M^2R^2$ are (3, -1, -1, -1, 0, 0, 0, 0) corresponding to $\Delta=(3, 1, 2)$.
   All the eight eigenvectors are given by
  \begin{eqnarray}
  u_1&=&(1, 1, 1, 1, 1, 1, 1, 1)\qquad u_2=(-1, 0, 0, 1, -1, 0, 0, 1)\nonumber\\
  u_3&=&(-1, 0, 1, 0, -1, 0, 1, 0)\qquad u_4=(-1, 1, 0, 0, -1, 1, 0, 0)\nonumber\\
  u_5&=&(0, 0, 0, -1, 0, 0, 0, 1)\qquad u_6=(0, 0, -1, 0, 0, 0, 1, 0)\nonumber\\
  u_7&=&(0, -1, 0, 0, 0, 1, 0, 0)\qquad u_8=(-1, 0, 0, 0, 1, 0, 0, 0).
  \end{eqnarray}
  As in the previous case, the flow ansatz is the combination $u_2+u_3+u_4$ which has eigenvalue -1, $\Delta=1$ and corresponds to a relevant operator.
  \item $f=-\frac{g_p}{g_n}$:
  \begin{equation}
  M^2=\frac{256 (g_n^2-g_p^2)^4}{(g_{2s}g_ng_p)^2}\left(
\begin{array}{cccccccc}
 \frac{3}{2} & 0 & 0 & 0 & \frac{3}{2} & 0 & 0 & 0 \\
 0 & \frac{3}{2} & 0 & 0 & 0 & -\frac{1}{2} & 1 & 1 \\
 0 & 0 & \frac{3}{2} & 0 & 0 & 1 & -\frac{1}{2} & 1 \\
 0 & 0 & 0 & \frac{3}{2} & 0 & 1 & 1 & -\frac{1}{2} \\
 \frac{3}{2} & 0 & 0 & 0 & \frac{3}{2} & 0 & 0 & 0 \\
 0 & -\frac{1}{2} & 1 & 1 & 0 & \frac{3}{2} & 0 & 0 \\
 0 & 1 & -\frac{1}{2} & 1 & 0 & 0 & \frac{3}{2} & 0 \\
 0 & 1 & 1 & -\frac{1}{2} & 0 & 0 & 0 & \frac{3}{2}
\end{array}
\right).
  \end{equation}
   The eigenvalues of $M^2R^2$ are (3, 3, 3, 3, 0, 0, 0, 0) corresponding to $\Delta=(3, 2)$.
   All the eight eigenvectors are given by
  \begin{eqnarray}
  w_1&=&(0, \frac{2}{3}, \frac{2}{3}, -\frac{1}{3}, 0, 0, 0, 1)\qquad w_2=(0, \frac{2}{3}, -\frac{1}{3}, \frac{2}{3}, 0, 0, 1, 0)\nonumber\\
  w_3&=&(0, -\frac{1}{3}, \frac{2}{3}, \frac{2}{3}, 0, 1, 0, 0)\qquad w_4=(1, 0, 0, 0, 1, 0, 0, 0)\nonumber\\
  w_5&=&(0, -\frac{2}{3}, -\frac{2}{3}, \frac{1}{3}, 0, 0, 0, 1)\qquad w_6=(0, -\frac{2}{3}, \frac{1}{3}, -\frac{2}{3}, 0, 0, 1, 0)\nonumber\\
  w_7&=&(0, \frac{1}{3}, -\frac{2}{3}, -\frac{2}{3}, 0, 1, 0, 0)\qquad w_8=(-1, 0, 0, 0, 1, 0, 0, 0).
  \end{eqnarray}
  Our flow corresponds to the combination $w_1+w_2+w_3-3w_4$ which has eigenvalue 3, $\Delta=3$. This is consistent with the fact that at the IR, the operator must be irrelevant.
\end{itemize}
We also compute the mass spectrum for the full scalar manifold.
Using gauge transformation, we are left with twenty scalars. At the
UV points $f=\pm1$, six of the extra twelve scalars have
$M^2R^2=-\frac{1}{4}$, and the other six are massless. At the IR
point $f=-\frac{g_p}{g_n}$, out of the extra twelve scalars, there
are six massless scalars and six scalars with $M^2R^2=\frac{3}{4}$.
\\ \indent
The behavior of the scalars at large $r$ is given by the linearized
equations
\begin{eqnarray}
\frac{da}{dr}&=&\frac{8 a_0}{b_0} [2 a(r) (2 a_0^2 (b_0^2
g_{2s}+g_n)+3 a_0 b_0 g_p+b_0^2 g_n)+b(r) (a_0^2 (b_0^2
g_{2s}-g_n)+b_0^2 g_n)\nonumber\\& &+a_0^2 b_0^2 g_{2s}+a_0^2 g_n+2
a_0 b_0
g_p+b_0^2 g_n]\nonumber\\
\frac{db}{dr}&=&\frac{8}{b_0}[3 a(r) (a_0^3 (b_0^2 g_{2s}-g_n)+a_0
b_0^2 g_n)+2 b_0^2 b(r) (a_0^3 g_{2s}+3 a_0 g_n+3 b_0
g_p)\nonumber\\& &+a_0^3 b_0^2 g_{2s}-a_0^3 g_n+3 a_0 b_0^2 g_n+2
b_0^3 g_p]
\end{eqnarray}
where $a_0$ and $b_0$ are the values of $a(r)$ and $b(r)$ at the
critical point. For the UV ($r\rightarrow\infty$) point, $f=1$ and
$f=-1$, we find
\begin{equation}
a(r)\sim e^{-r/R}, \qquad b(r)\sim e^{-r/R}.
\end{equation}
For the IR point ($r\rightarrow-\infty$), we find
\begin{equation}
a(r)\sim e^{r/R}, \qquad b(r)\sim e^{r/R}.
\end{equation}
The general behavior of a scalar field near the UV fixed point is
given by \cite{ps}, \cite{muck}
\begin{equation}
\phi(x,r)=e^{-(2-\Delta)r}(1+\ldots)\hat{\phi}(x)+e^{-\Delta
r}(1+\ldots)\check{\phi}(x)\label{scalar},
\end{equation}
where $\hat{\phi}(x)$ and $\check{\phi}(x)$ correspond to the source
and the vacuum expectation value of the operator of dimension
$\Delta$, respectively \cite{muck}, \cite{kw}. In (\ref{scalar}),
$1<\Delta\leq 2$. For $\Delta=1$ or $\Delta=\frac{d}{2}$ in $d$
dimensional field theory, the behavior of the scalar is given by
\cite{muck}
\begin{equation}
\phi(r,x)=e^{-r/R}\big(\frac{r}{R}\hat{\phi}(x)+\check{\phi}(x)\big)+\ldots.\label{scalar1}
\end{equation}
We see that in our flow, the first term in (\ref{scalar1}) is
absent, so there is no source. The flow is therefore of the
so-called v.e.v. type, corresponding to the deformation of the UV
theory by an expectation value of an operator of dimension one. Near
the IR point, the scalar behaves as $e^{(\Delta-2)r/R}$
\cite{freedman}. We then find that, in the IR, the corresponding
operator is irrelevant with dimension 3.
\subsection{The Flow Between $(2,0)$ Vacua}
\begin{figure}[!h] \centering
  \includegraphics[width=0.5\textwidth, bb = 0 0 200 150 ]{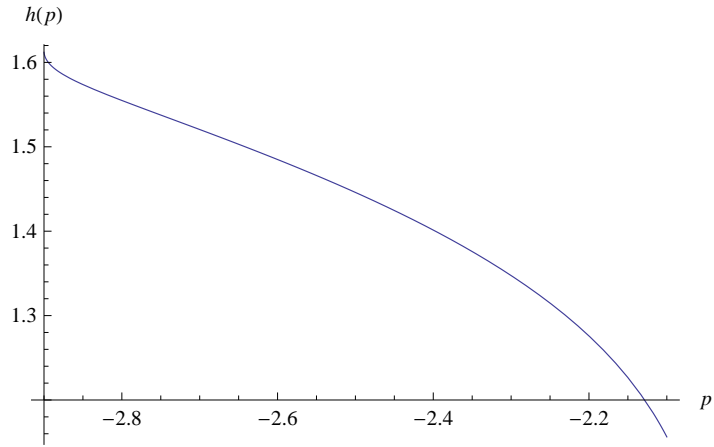}\\
  \caption{$h(p)$ solution.}
  \label{fig1}
\end{figure}

\begin{figure}[!h] \centering
  \includegraphics[width=0.5\textwidth, bb = 0 0 200 150 ]{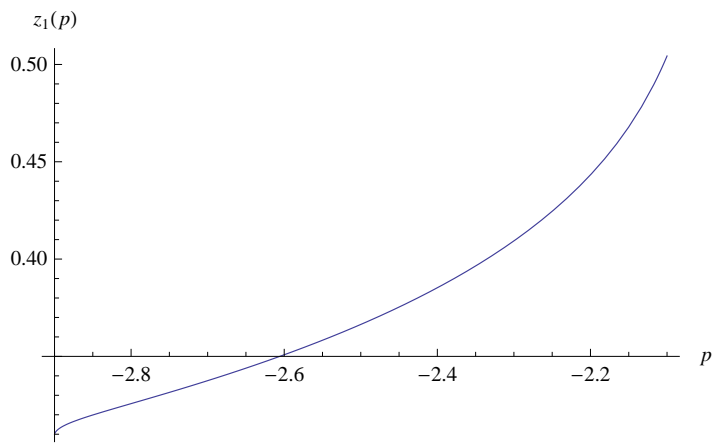}\\
  \caption{$z_1(p)$ solution.}
  \label{fig2}
\end{figure}

\begin{figure}[!h] \centering
  \includegraphics[width=0.5\textwidth, bb = 0 0 200 150 ]{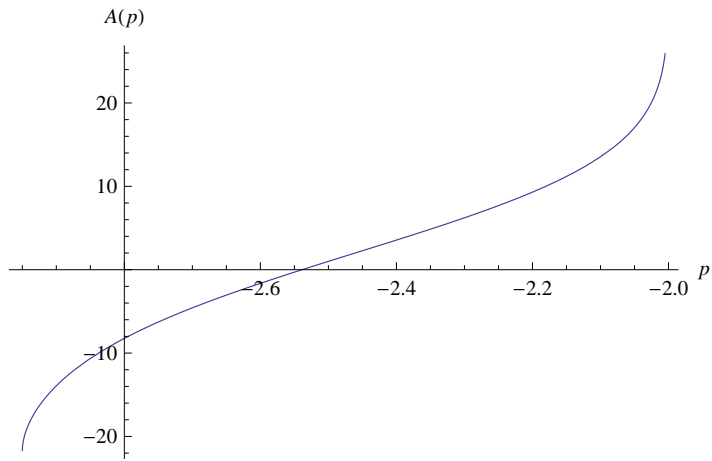}\\
  \caption{$A(p)$ solution.}
  \label{fig3}
\end{figure}

Now, we consider the flow between IV and V critical points.
\\ \indent We begin by giving the ansatz for $e_1$ and $e_2$,
\begin{eqnarray}
e_1&=&\sqrt{\frac{2(g_p-g_n)}{g_{2s}}}\left(
                                      \begin{array}{cccc}
                                        x(r) & 0 & 0 & 0 \\
                                        0 & q(r) & 0 & 0 \\
                                        0 & 0 & z(r) & 0 \\
                                        0 & 0 & 0 & z(r) \\
                                      \end{array}
                                    \right)\nonumber \\
e_2&=&\sqrt{\frac{2(g_p-g_n)}{g_{2s}}}\left(
                                      \begin{array}{cccc}
                                        y(r) & 0 & 0 & 0 \\
                                        0 & y(r) & 0 & 0 \\
                                        0 & 0 & w(r) & 0 \\
                                        0 & 0 & 0 & w(r) \\
                                      \end{array}
                                    \right).
\end{eqnarray}
Consistency condition for the BPS equations requires
\begin{eqnarray}
x&=&-\frac{(g_n+g_p)y^2}{q(g_n+g_p-2g_ny^2)}\\
w&=&\sqrt{\frac{g_n+g_p}{g_n+g_p+2g_pz^2}}z.
\end{eqnarray}
The $\delta\chi^{iI}$ equations give
\begin{eqnarray}
\frac{dz}{dr}&=&\frac{1}{g_{2s}(g_n+g_p)q^2y^2(g_n+g_p-2g_ny^2)}\{8(g_n+g_p)z^3(2q^2y^2(2(g_n^3-g_ng_p^2)y^4\nonumber\\&
&+(2g_p^3+6g_ng_p^2-4g_n^3)y^2 +(g_n-2g_p)
(g_n+g_p)^2)+g_nq^4(g_n+g_p-2g_ny^2)^2\nonumber\\&
&+g_n(g_n+g_p)^2y^4)\}\label{dz} \\
\frac{dy}{dr}&=&\frac{8y(g_n+g_p-2g_ny^2)}{g_{2s}(g_n+g_p)}\Bigg\{-\frac{2(g_n+g_p)y^2}{g_n+g_p}\bigg(
(g_n-g_p)^2z^2-2g_n^2+3g_ng_p-g_p^2\nonumber\\ &
&+\frac{(g_n-g_p)^2(g_n+g_p)z^2}{g_n+g_p+2g_pz^2}\bigg)+\frac{g_p(g_p-g_n)(g_n+g_p)^2y^4
}{q^2(g_n+g_p-2g_ny^2)^2}+g_p(g_p-g_n)q^2\Bigg\} \label{dy} \\
\frac{dq}{dr}&=&-\frac{8(g_n-g_p)q(g_n+g_p-2g_ny^2)}{g_{2s}(g_n+g_p)y^2}\Bigg\{\frac{(g_n+g_p)^2y^4}{q^2(g_n+g_p-2g_ny^2)}
\bigg(g_nz^2-g_py^2 \nonumber\\&
&+\frac{g_n(g_n+g_p)z^2}{g_n+g_p+2g_pz^2}\bigg)+q^2\bigg(\frac{2(g_n+g_p)^2y^4(g_p-2g_n+(g_n-g_p)z^2)}{q^2(g_n+g_p-2g_ny^2)^2}
\nonumber\\& &+\frac{(g_n+g_p)z^2}{g_n+g_p+2g_pz^2}
\big(\frac{2(g_n-g_p)(g_n+g_p)^2y^4}{q^2(g_n+g_p-2g_ny^2)}-g_n\big)-g_nz^2+g_py^2\bigg)
\nonumber\\&
&+\frac{2g_p(g_n+g_p)q^2y^2}{g_n+g_p-2g_ny^2}\Bigg\}.\label{dq}
\end{eqnarray}
This flow ansatz preserves (2,0) supersymmetry along the entire
flow. We now change the variables to $z_1$, $h$, and $p$
\begin{eqnarray}
y&=&\sqrt{\frac{g_n+g_p}{2g_n(1+z_1)}}\\
z&=&\sqrt{\frac{g_n+g_p}{2g_ph}}\\
q&=&\sqrt{-\frac{(g_n+g_p)\sqrt{p^2-4}}{g_nz_1(p^2-4+p\sqrt{p^2-4})}}
\end{eqnarray}
and rescale $r$ to $r \frac{8(g_n^2-g_p^2)}{g_{2s}g_ng_p}$. The
final forms of (\ref{dz}), (\ref{dy}), and (\ref{dq}) are
\begin{eqnarray}
\frac{dz_1}{dr}&=&\frac{(g_n^2-g_p^2-h(g_p^2p-2g_n(g_n-2g_p)+g_ph(4g_n-2g_p+g_pp)))}{h(h+1)}\\
\frac{dh}{dr}&=&\frac{g_n^2-g_p^2+z_1(g_n^2p(1+z_1)-2(g_p(g_p-2g_n)+g_n(g_n-2g_p)z_1))}{
z_1(1+z_1)}\\
\frac{dp}{dr}&=&-(p^2-4)\bigg[g_n^2\bigg(\frac{1}{h}+\frac{1}{1+h}\bigg)-\frac{g_p^2}{z_1}-\frac{g_p^2}{1+z_1}\bigg].
\end{eqnarray}
We proceed by taking $p$ as an independent variable and obtain
\begin{eqnarray}
\frac{dz_1}{dp}&=&\frac{(g_p^2-g_n^2+(g_p^2p+4g_ng_p-2g_n^2)h+g_p(4g_n+g_p(p-2))h^2)z_1(1+z_1)}{
(p^2-4)(g_n^2(1+2h)z_1(1+z_1)-g_p^2h(1+h)(1+2z_1))}\\
\frac{dh}{dp}&=&\frac{h(1+h)(g_p^2-g_n^2+2g_p(g_p-2g_n)z_1+2g_n(g_n-2g_p)z_1^2-g_n^2pz_1(1+z_1))}
{(p^2-4)(g_n^2(1+2h)z_1(1+z_1)-g_p^2h(1+h)(1+2z_1))}\nonumber\\.
\end{eqnarray}
Recall that $g_n=tg_p$, we find that the two critical points are
given by
\begin{itemize}
  \item IV: \begin{eqnarray}
  p&=&-2,\qquad h=\frac{1}{4}(t-1+\sqrt{5+2t+t^2}), \nonumber \\ & &
  \textrm{and}\qquad z_1=\frac{1-t+\sqrt{1+2t+5t^2}}{4t},
  \end{eqnarray}\\ and
  \item V: \begin{eqnarray}
  p&=&2-\frac{2}{t}-2t,\qquad h=\frac{1}{2}(t-1+\sqrt{1+t^2}), \nonumber \\ & &
  \textrm{and}\qquad z_1=\frac{1-t+\sqrt{1+t^2}}{2t}.
  \end{eqnarray}
\end{itemize}
We now give the numerical solution. Choosing $t=2$, we find the
numerical values for the critical points
\begin{eqnarray}
\textrm{IV}:\qquad p&=&-2.000,\qquad h=1.151,\qquad z_1=0.500 \nonumber\\
\textrm{V}:\qquad p&=&-3.000,\qquad h=1.618,\qquad z_1=0.309.
\end{eqnarray}
The numerical solutions for the flow are given in Fig.\ref{fig1} and
Fig.\ref{fig2}.
\\ \indent The gravitino variation gives an equation for $A(p)$, with $t=2$,
\begin{eqnarray}
\frac{dA}{dp}&=&-\frac{8 g_p^2 [(p^2-2)\sqrt{p^2-4}+p^3-4p]
}{\sqrt{p^2-4}
(p\sqrt{p^2-4}+p^2-4)^2}\times \nonumber\\
& & [(p+6) h(p)^2 (2 z_1(p)+1)+p h(p) (1-2 z_1(p) (4
z_1(p)+3))\nonumber\\
& &-2 z_1(p) (2 p z_1(p)+2 p+3)-3]/
[-4 (2 h(p)+1) z_1(p)^2\nonumber\\
& &+2 ((h(p)-3) h(p)-2)z_1(p)+h(p) (h(p)+1)].
\end{eqnarray}
\indent Choosing $g_{2s}=-1$ and $g_p=1$, we find the numerical
solution for $A$ as shown in Fig.\ref{fig3}.
\\ \indent In this flow, the point IV is the UV fixed point, and V is the IR. The ratio of the central charges is
\begin{equation}
\frac{c_{UV}}{c_{IR}}=\frac{(g_n+g_p)^2}{4g_ng_p}.
\end{equation}
This ratio is greater than 1 in consistent with the c-theorem. We
also compute the scalar mass matrices at both critical points, but
the form of the matrices is too complicated to be written here. We
give only the numerical values of the eigenvalues in our choice of
$g_p=1$, $g_n=2$ and $g_{2s}=-1$.
\begin{itemize}
  \item IV: Eigenvalues of $M^2R^2$ are $(3.70, -1.00, -1.00, -0.97, 0.36, 0.36, 0.00, 0.00)$\\
  with eigenvectors
  \begin{eqnarray}
  U_1&=&(-0.47, -0.47, -0.44, -0.44, -0.16, -0.16, -0.24, -0.24)\nonumber\\ U_2&=&(0.33, -0.33, 0.44, -0.44, 0.00, 0.00, 0.44, -0.44)\nonumber\\
  U_3&=&(0.63, -0.63, -0.23, 0.23, 0.00, 0.00, -0.23, 0.23)\nonumber\\ U_4&=&(0.47, 0.47, -0.44, -0.44, 0.16, 0.16, -0.24, -0.24)\nonumber\\
  U_5&=&(0.00, 0.00, -0.49, 0.49, -0.14, 0.14, 0.49, -0.49)\nonumber\\ U_6&=&(0.00, 0.00, -0.10, 0.10, 0.69, -0.69, 0.10, -0.10)\nonumber\\
  U_7&=&(0.22, 0.22, -0.06, -0.06, -0.66, -0.66, 0.11, 0.11)\nonumber\\ U_8&=&(-0.04, -0.04, -0.33, -0.33, 0.12, 0.12, 0.61, 0.61).
  \end{eqnarray}
  Our flow ansatz corresponds to $U_4$ with $\Delta=1.168$ which is dual to a relevant
  operator. Note also that, our ansatz does not correspond to the
  one which saturates the
  bound $M^2R^2=-1$. This means the dual operator is not the most
  relevant one.
  \item V: Eigenvalues of $M^2R^2$ are $(4.17, 3.33, 3.33, 3.33, 0.84, 0.84, 0.84, 0.00)$\\
   with eigenvectors
  \begin{eqnarray}
  V_1&=&(-0.211, -0.894, -0.211, -0.211, -0.130, -0.130, -0.130, -0.130)\nonumber\\ V_2&=&(0.201, -0.031, 0.063, -0.159, -0.609, 0.001, 0.090, 0.742)\nonumber\\
  V_3&=&(0.390, -0.398, 0.523, 0.432, 0.400, 0.011, -0.103, 0.241)\nonumber\\ V_4&=&(-0.293, 0.159, 0.015, -0.258, 0.359, -0.004, -0.801, 0.227)\nonumber\\
  V_5&=&(0.255, 0.002, -0.712, 0.572, -0.039, 0.086, -0.300, 0.046)\nonumber\\ V_6&=&(-0.146, 0.011, 0.387, 0.287, -0.526, 0.391, -0.384, -0.411)\nonumber\\
  V_7&=&(0.757, 0.004, -0.047, -0.499, 0.007, 0.156, -0.207, -0.328)\nonumber\\ V_8&=&(0.130, 0.130, 0.130, 0.130, -0.211, -0.893, -0.211, -0.211).
  \end{eqnarray}
  Our flow ansatz corresponds to $V_1$ with $\Delta=3.275$ which is dual to an irrelevant operator.
\end{itemize}
\indent The behavior near $r\rightarrow\infty$ can be obtained as in
the previous case. With $g_p=1$, $g_n=2$, and $g_{2s}=-1$, we find
that
\begin{eqnarray}
p(r) &\sim &  e^{-2r/R},\qquad z_1(r),\phantom{a} h(r)\sim
e^{-1.168r/R}.
\end{eqnarray}
At the IR point, we find
\begin{equation}
z_1(r),\phantom{a} h(r),\phantom{a} p(r)\sim e^{1.275r/R}.
\end{equation}
From the dominant term near the UV fixed point, we see that the flow
solution describes the deformation of the UV theory by a vacuum
expectation value of an operator of dimension $1.168$. We find that
this flow is also a v.e.v. flow. The corresponding operator in the
IR theory is an irrelevant operator of dimension $3.275$.
\section{Vacua of the $N=8$ Theory}
In this section, we study a gauging of an $N=8$ theory. We restrict
our discussion to the target space $\frac{SO(8,8)}{SO(8)\times
SO(8)}$. We parametrize the coset elements $L$ as in the $N=4$ case,
but now obviously
 $e$ is an element of $GL(8, \bf R)$ and $B$ is an antisymmetric
$8\times 8$  matrix. The resulting $L$ depends on 92 parameters,
but, again using the right action of a diagonal $SO(8)$, one can
bring $e$ to an upper triangular form, thereby reducing the number
of parameters to 64.  As for the non compact generators, the
$Y^{ab}$ introduced before carry over in the obvious way to the
present case, with $a,b=1,\dots,8$.\\ \indent We are going to gauge
the subgroup $(SO(4)\ltimes {\bf R}^6)^2$. Accordingly, we introduce
gauge group generators:
\begin{equation}
t^{\mathcal A}=\left(
      \begin{array}{cccc}
        a_1 & 0 & 0 & 0 \\
        0 & a_2 & 0 & 0 \\
        0 & 0 & a_1 & 0 \\
        0 & 0 & 0 & a_2 \\
      \end{array}
    \right),
\qquad t^{\mathcal{B}}=\left(
      \begin{array}{cccc}
        b_1 & 0 & b_1 & 0 \\
        0 & b_2 & 0 & b_2 \\
        -b_1& 0 & -b_1 & 0 \\
        0 & -b_2 & 0 & -b_2 \\
      \end{array}
    \right).
\end{equation}
Here all entries are $4\times 4$ matrices, $a_1$ $(a_2)$ are
generators of the first (second) $SO(4)$, $b_1$ and $b_2$ are
antisymmetric and correspond to independent shifts of $B$. More
precisely, the upper and lower $4\times 4$ diagonal blocks of $B$
will be shifted by $2b_1$ and $2b_2$, respectively, and therefore
could be set to zero. Generators carrying index 1 commute with those
carrying index 2, and one checks the structure of the gauge group
stated above. The $f$-tensors are constructed as follows: we choose
a basis of symmetric, real $SO(8)$ $\gamma$-matrices with $8\times
8$ off-diagonal blocks $\Gamma^I$, so that:
\begin{equation}
f^{IJ}_{ab,
cd}=-\frac{1}{2}\textrm{Tr}(\varepsilon^{ba}[\Gamma^I,\Gamma^J]\varepsilon^{cd}).
\end{equation}
As for the embedding tensor $\Theta$, the structure discussed in the
$N=4$ case extends naturally to the present case, and now we expect
a priori 8 couplings corresponding to the 8 $SU(2)$'s (including the
$\mathcal B$ generators).  We then proceed first by computing the
$\mathcal V$'s which are given in the Appendix \ref{formulae} and
then the $T$ tensors which are given by:
\begin{eqnarray}
T^{LJ, MK}&=&g_{1s}(\mathcal{V}^{LJ,
PQ}_{+\textrm{a}}\mathcal{V}^{MK,PQ}_{+\textrm{b}}+\mathcal{V}^{LJ,
PQ}_{+\textrm{b}}\mathcal{V}^{MK,PQ}_{+\textrm{a}})+g_{1a}(\mathcal{V}^{LJ,
PQ}_{-\textrm{a}}\mathcal{V}^{MK,PQ}_{-\textrm{b}}\nonumber\\
& &+\mathcal{V}^{LJ,
PQ}_{-\textrm{b}}\mathcal{V}^{MK,PQ}_{-\textrm{a}})+g_{2s}(\mathcal{V}^{LJ,
P'Q'}_{+\textrm{a}}\mathcal{V}^{MK,P'Q'}_{+\textrm{b}}+\mathcal{V}^{LJ,
P'Q'}_{+\textrm{b}}\mathcal{V}^{MK,P'Q'}_{+\textrm{a}})\nonumber\\
& &+g_{2a}(\mathcal{V}^{LJ,
P'Q'}_{-\textrm{a}}\mathcal{V}^{MK,P'Q'}_{-\textrm{b}}+\mathcal{V}^{LJ,
P'Q'}_{-\textrm{b}}\mathcal{V}^{MK,P'Q'}_{-\textrm{a}})+h_{1s}\mathcal{V}_{+\textrm{b}}^{LJ,PQ}\mathcal{V}_{+\textrm{b}}^{MK,PQ}\nonumber\\
&
&+h_{1a}\mathcal{V}_{-\textrm{b}}^{LJ,PQ}\mathcal{V}_{-\textrm{b}}^{MK,PQ}+h_{2s}\mathcal{V}_{+\textrm{b}}^{LJ,P'Q'}\mathcal{V}_{+\textrm{b}}^{MK,P'Q'}+
h_{2a}\mathcal{V}_{-\textrm{b}}^{LJ,P'Q'}\mathcal{V}_{-\textrm{b}}^{MK,P'Q'},
\nonumber \\
T^{LJ}_{ab}&=&g_{1s}(\mathcal{V}_{+\textrm{a}}^{LJ,PQ}{\mathcal{V}_{+\textrm{b}}}^{PQ}_{ab}+
\mathcal{V}_{+\textrm{b}}^{LJ,PQ}{\mathcal{V}_{+\textrm{a}}}^{PQ}_{ab})+g_{1a}(\mathcal{V}_{-\textrm{a}}^{LJ,PQ}{\mathcal{V}_{-\textrm{b}}}^{PQ}_{ab}\nonumber\\
& &+
\mathcal{V}_{-\textrm{b}}^{LJ,PQ}{\mathcal{V}_{-\textrm{a}}}^{PQ}_{ab})
+g_{2s}(\mathcal{V}_{+\textrm{a}}^{LJ,P'Q'}{\mathcal{V}_{+\textrm{b}}}^{P'Q'}_{ab}+
\mathcal{V}_{+\textrm{b}}^{LJ,P'Q'}{\mathcal{V}_{+\textrm{a}}}^{P'Q'}_{ab})\nonumber\\
&
&+g_{2a}(\mathcal{V}_{-\textrm{a}}^{LJ,P'Q'}{\mathcal{V}_{-\textrm{b}}}^{P'Q'}_{ab}
+\mathcal{V}_{-\textrm{b}}^{LJ,P'Q'}{\mathcal{V}_{-\textrm{a}}}^{P'Q'}_{ab})+h_{1s}\mathcal{V}^{LJ,PQ}_{+\textrm{b}}{\mathcal{V}_{+\textrm{b}}}^{PQ}_{ab}
\nonumber\\
&
&+h_{1a}\mathcal{V}^{LJ,PQ}_{-\textrm{b}}{\mathcal{V}_{-\textrm{b}}}^{PQ}_{ab}+h_{2s}\mathcal{V}^{LJ,P'Q'}_{+\textrm{b}}{\mathcal{V}_{+\textrm{b}}}^{P'Q'}_{ab}
+h_{2a}\mathcal{V}^{LJ,P'Q'}_{-\textrm{b}}{\mathcal{V}_{-\textrm{b}}}^{P'Q'}_{ab},
\end{eqnarray}
where $P, Q,\ldots=1,\ldots,4$ and $P', Q',\ldots=5,\ldots,8$. Here
$L,J,M,K$ are $SO(8)$ R-symmetry indices, and $a,b=1,\dots,8$ label
the 64 non-compact generators in $SO(8,8)$. $P,Q=1,\dots,4$ and
$P',Q'=5,\dots,8$  label the first and second $SO(4)$, respectively.
We have included also the 8 coupling constants, but actually,
consistency imposes relations among them:
\begin{eqnarray}
g_{1a}&=&-g_{1s},\qquad g_{2a}=-g_{2s}\nonumber\\
h_{1a}&=&-h_{1s},\qquad \textrm{and}\qquad h_{2a}=-h_{2s}.
\end{eqnarray}
Notice that if we set the type-2 couplings to zero i.e.
$g_{2s}=g_{2a}=h_{2s}=h_{2a}=0$, we decouple the second $SO(4)$ and
therefore we recover a truncation of the single $SO(4)$ gauging
studied in \cite{ns2} as the supergravity dual of the D1-D5 system
in IIB theory on $K3$ or $T^4$.  It can be obtained by reducing
(2,0) six-dimensional supergravity on $AdS_3\times S^3$.\\ \indent A
simple class of supersymmetric $AdS$ vacua can be obtained as
follows. We parameterize $e$ and $B$ as:
\begin{equation}
e=\left(
      \begin{array}{cccccccc}
        a_1 & 0 & 0 & 0 & e_{15} & e_{16} & e_{17} & e_{18} \\
        0 & a_2 & 0 & 0 & e_{25} & e_{26} & e_{27} & e_{28} \\
        0 & 0 & a_3 & 0 & e_{35} & e_{36} & e_{37} & e_{38} \\
        0 & 0 & 0 & a_4 & e_{45} & e_{46} & e_{47} & e_{48} \\
        0 & 0 & 0 & 0 & a_5 & 0 & 0 & 0 \\
        0 & 0 & 0 & 0 & 0 & a_6 & 0 & 0 \\
        0 & 0 & 0 & 0 & 0 & 0 & a_7 & 0 \\
        0 & 0 & 0 & 0 & 0 & 0 & 0 & a_8 \\
      \end{array}
    \right)
    \end{equation}
\begin{equation}
B=\left(
    \begin{array}{cccccccc}
      0 & 0 & 0 & 0 & b_{15} & b_{16} & b_{17} & b_{18} \\
      0 & 0 & 0 & 0 & b_{25} & b_{26} & b_{27} & b_{28} \\
      0 & 0 & 0 & 0 & b_{35} & b_{36} & b_{37} & b_{38} \\
      0 & 0 & 0 & 0 & b_{45} & b_{46} & b_{47} & b_{48} \\
      -b_{15} & -b_{25} & -b_{35} & -b_{45} & 0 & 0 & 0 & 0 \\
      -b_{16} & -b_{26} & -b_{36} & -b_{46} & 0 & 0 & 0 &0 \\
      -b_{17} & -b_{27} & -b_{37} & -b_{47} & 0 & 0 & 0 & 0 \\
      -b_{18} & -b_{28} & -b_{38} & -b_{48} & 0 & 0 & 0& 0 \\
      \end{array}
  \right).
\end{equation}
We have used the shift symmetry to set to zero the diagonal $4\times
4$ blocks of $B$ and the $SO(4)\times SO(4)$ left action to
diagonalize the diagonal blocks of $e$. For diagonal $e=(a_1, a_2,
a_3, a_4, a_5, a_6, a_7, a_8)$ and $B=0$, we cannot find any
interesting solutions apart from the trivial one with (4,4)
supersymmetry. All the truncations below have been checked to be
consistent, in the sense that there are no tadpoles for the
remaining scalars.
\\ \indent We find a class of
solutions by setting:
\begin{eqnarray}
a_2&=&a_3=a_4=a_1\nonumber\\
a_6&=&a_7=a_8=a_5\nonumber\\
b_{15}&=&\frac{1}{4}(c_{15}-c_{26}-c_{37}+c_{48})\qquad
b_{16}=\frac{1}{4}(-c_{16}-c_{25}-c_{38}-c_{47})\nonumber\\
b_{17}&=&\frac{1}{4}(c_{18}+c_{27}-c_{36}-c_{45})\qquad
b_{18}=\frac{1}{4}(c_{17}-c_{28}+c_{35}-c_{46})\nonumber\\
b_{25}&=&\frac{1}{4}(-c_{16}-c_{25}+c_{38}+c_{47})\qquad
b_{26}=\frac{1}{4}(-c_{15}+c_{26}-c_{37}+c_{48})\nonumber\\
b_{27}&=&\frac{1}{4}(c_{17}-c_{28}-c_{35}+c_{46})\qquad
b_{28}=\frac{1}{4}(-c_{18}-c_{27}-c_{36}-c_{45})\nonumber\\
b_{35}&=&\frac{1}{4}(c_{18}-c_{27}+c_{36}-c_{45})\qquad
b_{36}=\frac{1}{4}(-c_{17}-c_{28}+c_{35}+c_{46})\nonumber\\
b_{37}&=&\frac{1}{4}(-c_{15}-c_{26}-c_{37}-c_{48})\qquad
b_{38}=\frac{1}{4}(-c_{16}+c_{25}+c_{38}-c_{47})\nonumber\\
b_{45}&=&\frac{1}{4}(-c_{17}-c_{28}-c_{35}-c_{46})\qquad
b_{46}=\frac{1}{4}(-c_{18}+c_{27}+c_{36}-c_{45})\nonumber\\
b_{47}&=&\frac{1}{4}(-c_{16}+c_{25}-c_{38}+c_{47})\qquad
b_{48}=\frac{1}{4}(c_{15}+c_{26}-c_{37}-c_{48}),
\end{eqnarray}
and all other parameters are zero. We can choose
\begin{eqnarray}
c_{16}&=&c_{17}=c_{18}=c_{25}=c_{27}=c_{28}=0\nonumber\\
c_{35}&=&c_{36}=c_{38}=c_{45}=c_{46}=c_{47}=0.
\end{eqnarray}
Supersymmetric vacua require
\begin{equation}
g_{1s}=-a_1^2h_{1s}\qquad g_{2s}=-a_5^2h_{2s}\qquad
h_{2s}=\frac{a_1^4}{a_5^4}h_{1s}.
\end{equation}
\begin{itemize}
  \item \textbf{(1,1) critical point}\\
  This point is given by $c_{15}=0$,\\
  \begin{eqnarray}
  A_1&=&(-\frac{16g_{1s}^2}{h_{1s}}, \frac{16g_{1s}^2}{h_{1s}},
-\frac{8g_{1s}^2}{h_{1s}}\sqrt{4+a_1^2a_5^2c_{26}^2},
          \frac{8g_{1s}^2}{h_{1s}}\sqrt{4+a_1^2a_5^2c_{26}^2},\nonumber\\& & -\frac{8g_{1s}^2}{h_{1s}}\sqrt{4+a_1^2a_5^2c_{37}^2},
          \frac{8g_{1s}^2}{h_{1s}}\sqrt{4+a_1^2a_5^2c_{37}^2}, -\frac{8g_{1s}^2}{h_{1s}}\sqrt{4+a_1^2a_5^2c_{48}^2},
          \nonumber\\& &\frac{8g_{1s}^2}{h_{1s}}\sqrt{4+a_1^2a_5^2c_{48}^2}
      )\label{n8a1}
  \end{eqnarray}
  and
  $V_0=-\frac{1024g_{1s}^4}{h_{1s}^2}$.
  \item \textbf{(2,2) critical point}\\
  This point is given by $c_{15}=0$ and $c_{26}=0$.
  \item \textbf{(3,3) critical point}\\
  This point is given by $c_{15}=0$,$c_{26}=0$ and $c_{37}=0$.
  \item \textbf{(4,4) critical point}\\
  This point is given by $c_{15}=0$, $c_{26}=0$, $c_{37}=0$ and $c_{48}=0$.
\end{itemize}
All of them have the same cosmological constant. $A_1$ for the last
three points is given by setting some of the appropriate values of
$c$'s to zero in (\ref{n8a1}). \\ \indent We also find other
solutions with non zero parameters
\begin{eqnarray}
& &a_2=a_3=a_4=a_1\qquad a_6=a_7=a_8=a_5\nonumber\\
& &e_{15}=e_{26}=e_{37}=e_{48}=e \nonumber\\
& &b_{16}=-b_{25}\qquad b_{38}=-b_{47}
\end{eqnarray}
subject to these relations $a_5^2+e^2=-\frac{g_{2s}}{h_{2s}}$,
$a_1^2=-\frac{g_{1s}}{h_{1s}}$ and
$\frac{g_{1s}^2}{h_{1s}}=\frac{g_{2s}^2}{h_{2s}}$. Note that in this
case, we also turn on some off-diagonal elements of $e$. The
solutions are given by:
\begin{itemize}
  \item \textbf{(2,2) critical point}\\
  This solution has $A_1=\frac{16g_{1s}^2}{h_{1s}}$ giving the same
cosmological constant as in the previous case.
  \item \textbf{(3,2) critical point}\\
  This can be obtained from the previous case by setting $b_{25}=b_{47}$ or
  $b_{25}=-b_{47}$.
\end{itemize}
So, there is no possible flow solution between all these critical
points.
\section{Conclusions}
In this paper, we have studied three dimensional gauged
supergravities and their $AdS_3$ supersymmetric vacua. We have
discussed the $N=4$ and $N=8$ theories with $SO(4) \ltimes \bf{R}^6$
and $(SO(4) \ltimes \bf{R}^6)^2$ gaugings, respectively. Several
supersymmetric $AdS_3$
vacua with different amount of supersymmetry have been found. \\
\indent In the $N=4$ theory, we have found analytic solutions
interpolating between two (3,1) vacua. These solutions describe
Renormalization Group flows between two fixed points of the dual
boundary field theory. We have checked  that the flows agree with
the c-theorem, in particular the central charges of UV fixed points
are strictly greater than those of the IR ones.  We have also found
a numerical solution describing the flow between (2,0) vacua with
similar qualitative features.  In both cases, we found  v.e.v.
flows, i.e. flows driven by vacuum expectation values of relevant
operators with dimensions $\Delta=1$ and $\Delta=1.168$,
respectively, as opposed to the most common case where the flow is
driven by a perturbing relevant operator.\\ \indent In the $N=8$
theory, we have found several vacua. However, they all have the same
cosmological constant/central charge and the flow issue does not
arise.
\\ \indent The gaugings considered here are of
non semi-simple Chern-Simons type, giving rise to semi-simple
Yang-Mills theories. In the $N=8$ case, the $(4,4)$ point is related
to the KK reduction of type IIB theory on $AdS_3\times S^3 \times
S^3\times S^1$, and it would be interesting to identify the marginal
deformations which take the theory to other less supersymmetric
vacua, i.e. to generalize the discussion of \cite{HS2}, where the
marginal deformation from $(4,4)$ to $(3,3)$ vacua has been worked
out in detail, to the $(k,k)$ vacua with $k<3$. The $N=4$ case seems
to be related, via a $\mathbb{Z}_2$ projection, to the $N=8$ theory,
and it would be interesting to see how this is acting on the
corresponding type IIB theory background. This would presumably help
us in understanding the nature of the dual $SCFT_2$.
\acknowledgments This work has been supported in part by the EU
grant UNILHC-Grant Agreement PITN-GA-2009-237920.
\appendix
\section{Essential formulae}\label{formulae}
In this appendix, we give the expressions for the $\mathcal V$'s.
Indices referring to each target space coordinates, $i, j, k,
\ldots$, will be traded by a pair of indices of the type $a, b,
c,\ldots$ from 1 to 4. Antisymmetric pairs of capital letters $I, J,
K, \ldots$ label $SO(4)$ adjoint indices.
\begin{eqnarray}
\mathcal{V}_{\pm
\textrm{a}}^{LJ,MK}&=&-\frac{1}{4}\textrm{Tr}[(e_1^tJ_+^{LJ}X_1^t+X_1J_+^{LJ}e_1)J_\pm^{MK}+(e_2^tJ_-^{LJ}X_2^t+X_2J_-
^{LJ}e_2)J_\pm^{MK}],\nonumber\\
{\mathcal{V}_{\pm 1,2
\textrm{a}}}^{MK}_{ab}&=&\textrm{Tr}[(e_{1,2}^t\varepsilon_{ab}X_{1,2}^t+
Y_{1,2}\varepsilon_{ab}e_{1,2})J_\pm^{MK}],\nonumber\\
\mathcal{V}_{\pm \textrm{b}}^{LJ,MK}&=&-\frac{1}{4}\textrm{Tr}[(e_1^tJ_+^{LJ}e_1^t+e_2^tJ_-^{LJ}e_2^t)J_\pm^{MK}],\nonumber\\
{\mathcal{V}_{\pm 1,2
\textrm{b}}}^{MK}_{ab}&=&\textrm{Tr}(e_{1,2}^t\varepsilon_{ab}e_{1,2}J_\pm^{MK}).
\end{eqnarray}
The string of indices $\pm 1,2 \textrm{a}$ ($\pm 1,2 \textrm{b}$)
indicates $\mathcal A$ ($\mathcal B$)-type gauging in the first
(second) space with (anti-)self-dual $SU(2)$. \\ \indent For
completeness, we give below the analogous expressions for the $N=8$
case:
\begin{eqnarray}
{\mathcal{V}_{{\pm \textrm {a}}}}^{LJ,MK}&=&\frac{1}{4\sqrt{2}}\textrm{Tr}[\Gamma^{JL}(eJ_\pm^{MK}X+X^tJ_\pm^{MK}e^t)],\nonumber\\
{\mathcal{V}_{\pm \textrm{b}}}^{LJ,MK}&=&\frac{1}{2\sqrt{2}} \textrm{Tr}[J_\pm^{JL}e\Gamma^{MK}e^t], \nonumber\\
{\mathcal{V}_{\pm\textrm{a}}}^{MK}_{ab}&=&\frac{1}{\sqrt{2}}\textrm{Tr}[\varepsilon_{ab}(X^t
J_\pm^{MK} e^t +eJ_\pm^{MK}Y)],
\nonumber\\
{\mathcal{V}_{\pm \textrm{
b}}}^{MK}_{ab}&=&\frac{2}{\sqrt{2}}\textrm{Tr}[\varepsilon_{ab}eJ_\pm^{MK}e^t].
\end{eqnarray}
Here $\Gamma^{JL}=-[\Gamma^J,\Gamma^L]/2$ and all indices run from 1
to 8 and $J_\pm^{MK}$ are the (anti-)self-dual $SU(2)$ generators in
$SO(4)\times SO(4)\subset SO(8)$, corresponding to the first
(second) $SO(4)$ for $M,K =1,\dots, 4$ ($M,K=5,\dots, 8$),
respectively.
\section{The other vacua of the $N=4$ theory}\label{vacua}
In this appendix, we give all vacua we have found in $N=4$ theory
apart from those involved in the flows.
\subsection{(4,0) vacuum}
\begin{itemize}
  \item VI. \begin{eqnarray}
  e_1&=&\sqrt{\frac{-2(g_n+g_p)}{g_{2s}}} \mathbb{I}_{4\times 4},\qquad e_2=\sqrt{\frac{2(g_p-g_n)}{g_{2s}}} \mathbb{I}_{4\times
  4},\nonumber\\ A_1&=&\frac{32g_ng_p}{g_{2s}},\qquad \textrm{and}\qquad
  V_0=-\frac{4096g_n^2g_p^2}{g_{2s}^2}.
\end{eqnarray}
\end{itemize}
\subsection{(3,0) vacua}
\begin{itemize}
            \item VII.
            \begin{eqnarray}e_1&=&a\Big(1,-\frac{g_m+g_p}{g_n+g_p+g_{2s}a^2},1,1\Big),\qquad
e_2=\sqrt{\frac{(g_p^2-g_n^2)}{g_p^2-g_n^2+g_{2s}g_pa}}a\mathbb{I}_{4\times 4},\nonumber\\
a&=&\sqrt{\frac{g_n^3-g_n^2g_p-g_ng_p^2+g_p^3+\sqrt{g_n^6-g_n^4g_p^2-g_n^2g_p^4+g_p^6}}{g_ng_pg_{2s}}}\nonumber\\
A_1&=&-\frac{8(g_n^2-g_p^2)^2}{g_{2s}g_ng_p},\qquad\textrm{and}\qquad
V_0=\frac{-256(g_n^2-g_p^2)^4}{(g_{2s}^2g_ng_p)^2}
\end{eqnarray}
            \item VIII.
            \begin{eqnarray}e_1&=&a\Big(1,-\frac{g_m+g_p}{g_n+g_p+g_{2s}a^2},1,1\Big),\qquad
e_2=\sqrt{\frac{(g_p^2-g_n^2)}{g_p^2-g_n^2+g_{2s}g_pa}}a\mathbb{I}_{4\times 4},\nonumber\\
a&=&\sqrt{\frac{g_n^3-g_n^2g_p-g_ng_p^2+g_p^3-\sqrt{g_n^6-g_n^4g_p^2-g_n^2g_p^4+g_p^6}}{g_ng_pg_{2s}}}\nonumber\\
A_1&=&-\frac{8(g_n^2-g_p^2)^2}{g_{2s}g_ng_p},\qquad\textrm{and}\qquad
V_0=\frac{-256(g_n^2-g_p^2)^4}{(g_{2s}^2g_ng_p)^2}
\end{eqnarray}
\end{itemize}
\subsection{(2,0) vacua}
\begin{itemize}
  \item IX.
  \begin{equation}
  e_1=-(a_1, a_1, b_1, b_2)\qquad e_2=(b_1, b_1, b_2, b_2)
  \end{equation}
  \begin{eqnarray}
a_1&=&2\sqrt{\frac{g_n^2-g_p^2}{g_{2s}(g_n-g_p+\sqrt{5g_n^2+2g_ng_p+g_p^2})}}\nonumber\\
a_2&=&2\sqrt{\frac{g_n^2-g_p^2}{g_{2s}(g_p-g_n+\sqrt{5g_p^2+2g_ng_p+g_n^2})}}\nonumber\\
b_1&=&2\sqrt{\frac{g_p^2-g_n^2}{g_{2s}(3g_n+g_p-\sqrt{5g_n^2+2g_pg_n+g_p^2})}}\nonumber\\
b_2&=&2\sqrt{\frac{g_n^2-g_p^2}{g_{2s}(\sqrt{g_n^2+2g_ng_p+5g_p^2}-g_n-3g_p)}}
 \end{eqnarray}
  \begin{eqnarray}
A_1&=&\frac{-32(g_n-g_p)^2}{g_{2s}}\qquad\textrm{and}\qquad
V_0=-\frac{4096(g_n-g_p)^4}{g_{2s}^2}.
\end{eqnarray}
  \item X.
  \begin{equation}
  e_1=(-a_1, a_1, a_2, a_2)\qquad e_2=(b_1, b_1, b_2, b_2)
 \end{equation}
 \begin{eqnarray}
   a_1&=&2\sqrt{\frac{g_p^2-g_n^2}{g_{2s}(g_p-g_n+\sqrt{5g_n^2+2g_pg_n+g_p^2})}}\nonumber\\
a_2&=&2\sqrt{\frac{g_p^2-g_n^2}{g_{2s}(g_n-g_p-\sqrt{5g_p^2+2g_pg_n+g_n^2})}}\nonumber\\
b_1&=&2\sqrt{\frac{g_p^2-g_n^2}{g_{2s}(3g_n+g_p+\sqrt{5g_n^2+2g_pg_n+g_p^2})}}\nonumber\\
b_2&=&2\sqrt{\frac{g_n^2-g_p^2}{g_{2s}(\sqrt{g_n^2+2g_ng_p+5g_p^2}-g_n-3g_p)}}
\end{eqnarray}
  \begin{eqnarray} A_1&=&\frac{-32(g_n-g_p)^2}{g_{2s}}\qquad\textrm{and}\qquad
V_0=-\frac{4096(g_n-g_p)^4}{g_{2s}^2}.
\end{eqnarray}
  \item XI.
  \begin{equation}
  e_1=(-a_1, a_1, a_2, a_2)\qquad e_2=(b_1, b_1, b_2, b_2)
 \end{equation}
\begin{eqnarray}
   a_1&=&2\sqrt{\frac{g_p^2-g_n^2}{g_{2s}(g_p-g_n-\sqrt{5g_n^2+2g_pg_n+g_p^2})}}\nonumber\\
a_2&=&2\sqrt{\frac{g_p^2-g_n^2}{g_{2s}(g_n-g_p+\sqrt{5g_p^2+2g_pg_n+g_n^2})}}\nonumber\\
b_1&=&2\sqrt{\frac{g_p^2-g_n^2}{g_{2s}(3g_n+g_p-\sqrt{5g_n^2+2g_pg_n+g_p^2})}}\nonumber\\
b_2&=&2\sqrt{\frac{g_p^2-g_n^2}{g_{2s}(\sqrt{g_n^2+2g_ng_p+5g_p^2}+g_n+3g_p)}}
\end{eqnarray}
  \begin{eqnarray} A_1&=&\frac{-32(g_n-g_p)^2}{g_{2s}}\qquad\textrm{and}\qquad
V_0=-\frac{4096(g_n-g_p)^4}{g_{2s}^2}.
\end{eqnarray}
\item XII.
\begin{equation}
  e_1=(-a_1, a_1, a_2, a_2)\qquad e_2=(b_1, b_1, b_2, b_2)
 \end{equation}
\begin{eqnarray}
   a_1&=&\sqrt{\frac{g_p-g_n}{g_{2s}}}
\sqrt{\frac{g_n-g_p+\sqrt{5 g_n^2+2 g_n
g_p+g_p^2}}{g_n}}\nonumber\\
a_2&=&\sqrt{\frac{g_p-g_n}{g_{2s}}}
\sqrt{-\frac{g_n-g_p+\sqrt{g_n^2+2 g_n g_p+5 g_p^2}}{g_p}}\nonumber\\
b_1&=&\sqrt{\frac{g_p-g_n}{g_{2s}}} \sqrt{\frac{3 g_n+g_p
  -\sqrt{5 g_n^2+2g_n g_p+g_p^2}}{g_n}}\nonumber\\
b_2&=&\sqrt{\frac{g_p^2-g_n^2}{g_{2s}g_p}}
\sqrt{-\frac{g_n-g_p+\sqrt{g_n^2+2 g_n g_p+
  5 g_p^2}}{2 g_p-\sqrt{g_n^2+2 g_n g_p+5 g_p^2}}}
\end{eqnarray}
  \begin{eqnarray} A_1&=&\frac{-32(g_n-g_p)^2}{g_{2s}}\qquad\textrm{and}\qquad
V_0=-\frac{4096(g_n-g_p)^4}{g_{2s}^2}.
\end{eqnarray}
\end{itemize}

\end{document}